\documentclass[prd,aps,nofootinbib,nobibnotes,superscriptaddress,preprintnumbers,11pt]{revtex4}

\usepackage{color}
\usepackage{amsmath} 
\usepackage{amsfonts}
\usepackage{graphicx}
\usepackage{slashed}
\usepackage[dvipsnames]{xcolor}
\definecolor{refs}{RGB}{245,156,74}

\usepackage{subcaption}
\usepackage{caption}
\usepackage{float}

\usepackage{physics}
\usepackage[colorlinks=true,hyperfootnotes=true,citecolor=cyan]{hyperref}

\newcommand{\be}{\begin{equation}}
\newcommand{\ee}{\end{equation}}

\definecolor{tabblue}{HTML}{1f77b4}
\definecolor{taborange}{HTML}{ff7f0e}
\definecolor{tabgreen}{HTML}{2ca02c}
\definecolor{tabred}{HTML}{d62728}
\definecolor{tabpurple}{HTML}{9467bd}

\begin{document}

\title{Matter influence  on large-scale scalar  dynamics}

\author{Philippe Brax}
\affiliation{Institut de Physique Th\'eorique, Universit\'e Paris-Saclay,CEA, CNRS, F-91191 Gif-sur-Yvette Cedex, France.}

\begin{abstract}
    Local structures in the Universe can influence the dynamics of light scalar fields when coupled to matter. We focus on light test fields evolving in matter modelled as a stochastic source. We describe the effective field theory for light scalars on large scales after integrating out the short distance dynamics. This is most conveniently performed in the Schwinger-Keldysh formalism where  we find that the large distance theory involves a stochastic noise corresponding to the exchanges between short and large scales, and new interactions which can affect the time evolution of light scalars. We exemplify this back-reaction when the coupling between matter and scalars is small leading to corrections to the Klein-Gordon equation of the light scalars on large scales. In particular, the resulting corrections to the scalar potential could lead to effects akin to dynamical dark energy.  We also consider the situation where all the substructures of the Universe are screened leading to the suppression of large scale dynamics and a cosmic Meissner effect. This highlights the potentially relevant effects of small scale structures on the cosmological dynamics of light scalar fields. 
\end{abstract}

\maketitle

\section{Introduction}

Light scalar fields may be at the origin of dynamical dark energy \cite{Copeland:2006wr,Joyce:2014kja,Brax:2017idh} as hinted by the DESI experiment \cite{DESI:2024mwx,DESI:2025fii, DESI:2025zgx,DES:2026jmi}. Unless protected by a symmetry, these scalars must couple to matter, whether Cold Dark Matter (CDM) or baryons \cite{Amendola:1999er,Brax:2025ahm}. When this is the case, there are strong observational constraints on the coupling constant $\beta$  between matter and the light scalars. Typically the Cassini bound \cite{Bertotti:2003rm} imposes that $\beta^2 \lesssim 10^{-5}$ when coupled to baryons whilst the Cosmic Microwave Background (CMB) alone imposes that $\beta\lesssim 10^{-2}$ for CDM only \cite{Archidiacono:2022iuu,Gomez-Valent:2020mqn}. 
The role of the coupling between matter and light scalars has become even more prominent recently as it offers an easy explanation to the apparent phantom crossing of the equation of state for dark energy \cite{DESI:2024mwx,DESI:2025fii, DESI:2025zgx,Das:2005yj,Wolf:2024stt,Khoury:2025txd}. Coupled models drive this behaviour at low redshift \cite{Martin:2005bp}. 

The coupling between scalars and matter does not have an influence on the large scale dynamics of the Universe only \cite{Brax:2021wcv,Burrage:2017qrf}. The behaviour of light scalars in local environments from the solar system to galactic scales is also strongly influenced by their coupling strength to matter. This is particularly the case when screening takes place \cite{Khoury:2003aq,Khoury:2003rn,Hinterbichler:2010es,Brax:2010gi}, i.e. the local behaviour of light scalars in the Solar system for instance will differ from the cosmological dynamics. This can be so different that gravitational tests in the solar system would be satisfied despite having a large coupling between matter and scalars on large scales \cite{Mota:2006ed}. As structures of widely different scales exist in the Universe,  the large scale dynamics of light scalars could be influenced by the local distribution of matter. This type of influence of small scales on large scales is common in particle physics where short distance physics leads to effective interactions for large distance fields when small distance physics is integrated out \cite{Burgess:2020tbq}. In this paper, we will consider the effects of integrating out the local dynamics of the Universe associated to  its matter distribution and obtain the large distance effective field theory for the light scalar fields. Typically, scales below a few Mpc's are integrated out. This will be obtained when the light scalars are test fields, i.e. they only respond to the dynamics of the Universe without influencing its course. This is not realistic when  the large scale dynamics of dark energy  influence the expansion of the Universe and cosmological perturbations which in turn would back-react of the growth of structure. This complex interplay is left for future work. Here we focus on a given matter distribution as emerging from the growth of structure in standard cosmology \cite{Huterer:2022dds}. 

The process of integrating out short scales is carried out in the Schwinger-Keldysh formalism \cite{Schwinger:1960qe,Keldysh:1964ud}. This is particularly useful as this allows one to keep track of the loss of information due to the integration process.  The system on large scale becomes open with exchanges between short and large scales \cite{Colas:2025ind}. This is captured by an effective noise representing the continuous exchange of information between scales. Technically, integrating out short scales is rendered more complex by the random nature of the matter distribution. For a given realisation of the Universe, the integration process can be performed in a traditional way. The fact that the matter distribution is stochastic implies that on cells of sizes the scale below which short distance effects are integrated out, the effective large scale dynamics will fluctuate from one cell to another across the Universe. These fluctuations correspond to different realisations of the matter distribution by ergodicity. We will average over the matter fluctuations to obtain the behaviour of the scalar field on large scales. This will correspond to the spatial average behaviour of the scalar field. 

We will analyse the large scale effective behaviour of the light scalar fields in two typical situations. The first one is the weak coupling limit where structures perturb the large scale dynamics perturbatively. In this case, we find that the Klein-Gordon equation governing the scalar dynamics on large scales receives corrections from induced operators in the effective action expressed in perturbation in the coupling constant $\beta$. This could lead to a potential  effect akin to dark energy \cite{Contaldi:2026vsw} and a novel way of tackling the coincidence problem.  At lowest order, the scalar perturbations on large scales obey the linear response theory. On the other hand, when all the structures of the Universe are screened, the resulting dynamics on large scales are tightly determined to be that of a rapidly oscillating scalar  at the background level and suppressed fluctuations with a propagation length  given by the small Compton wavelength of the scalar \cite{Khoury:2003aq,Khoury:2003rn}. This is the analogue of a cosmological Meissner effect.

In realistic situations where the light scalars would partake in dark energy, major differences would ensue. First of all, the growth of matter would be influenced by dark energy due to the coupling to matter $\beta$ and then back-react on the large scale dynamics of dark energy. Contrary to the case treated in this paper, not all substructures would be screened as can be seen on screening maps of the Universe obtained via N-body simulations \cite{Shao:2019wit}. Both effects would alter the overall picture for a test field presented here. The analysis of this more realistic case is left for future work.

The paper is arranged as follows. In section \ref{sec:kel}, we present the Schwinger-Keldysh formalism applied to large scales. Then in section \ref{sec:wil} we develop the Wilsonian point of view where short scales are integrated out. In section \ref{sec:grand} we focus on the exchange of information between short and large scales, define the statistical properties of the resulting noise term and deduce the properties of the Klein-Gordon equation for large scales. In section \ref{sec:pert}, we consider perturbations to the background evolution in the unscreened and completely screened cases. One appendix summarises known properties of the Schwinger-Keldysh formalism. Another one recalls the analogy with photons in matter. 

\section{The Schwinger-Keldysh description}
\label{sec:kel}

\subsection{The matter distribution} 
 
We are interested in the interactions between a self-interacting scalar field and matter when the matter distribution is random. This applies to astrophysical situations where we will assume that the matter distribution is Gaussian and characterised by its power spectrum 
\be
\langle e^{-i\int d^4x \sqrt{-g} j(x)S(x)}\rangle = e^{-\frac{1}{2}\int d^4 x d^4 y \sqrt{-g}_x \sqrt{-g}_y S(x)P(x;y)S(y)} \, ,
\ee
where $P$ is the two-point correlation function defined as
\be 
\langle j(x) j(y) \rangle= P(x;y) \, ,
\ee
where $P$ is a symmetric function of its arguments. 
\subsubsection{Time-local and translation invariant processes}

A simplifying and unrealistic assumption corresponds to taking that time and space translation invariance are respected by the matter distribution. This is useful as an illustration compared to the full complexity of the matter power spectrum deduced from the growth of cosmological perturbations.  This hypothesis implies that
\be 
P(x;y)= P(\vec x -\vec y, t-t') \, ,
\ee
where $x=(\vec x,t)$ and $y=(\vec y, t')$ are the 4-four vectors defining space-time positions. One can further simplify the treatment by imposing a Markovian ansatz for the correlator 
\be 
\label{timelocal}
P(\vec x-\vec y, t-t')= P(\vec x-\vec y) \delta (t-t') \, ,
\ee
which corresponds to taking the correlation time of the matter distribution to be much shorter than the characteristic timescales of the scalar dynamics. In a realistic cosmological context this approximation is not valid, and in the following section we will loosen this assumption. The spatial dependence is described by the correlator $P(\vec x - \vec y)$, whose Fourier transform 
\be 
P(\vec x - \vec y)= \int \frac{d^3 k}{(2\pi)^3}  P(\vec k) e^{i\vec k \cdot (\vec x - \vec y)} \, ,
\ee
defines the corresponding power spectrum $P(\vec k)$. In the general case we can  decompose the random field itself as
\be 
j(\vec x, t)= \int \frac{d^4k}{(2\pi)^4} j(\vec k, \omega) e^{ik.x} \, ,
\ee
where $k\cdot x= -\omega t + \vec k\cdot \vec x$. From this, we can read the correlator in momentum space, as\footnote{ We define $\slashed{dk}= \frac{dk}{2\pi}$ and $\slashed{\delta}(k)= 2\pi \delta (k)$.}
\be 
\langle j(\vec k, \omega) j(\vec k',\omega')= P(\vec k,\omega) \slashed{\delta}^{(3)}(\vec k + \vec k')\slashed{\delta }(\omega +\omega') \, .
\ee
Isotropy implies that $P(\vec k, \omega)$ depends on $k \equiv \vert \vec k \vert$, as well as
\be 
P(k, -\omega)= P^\dagger(k, \omega) \, .
\ee
Under the Markovian ansatz (\ref{timelocal}), translational invariance and the fact that matter densities are always real, imply that $P(k)$ is independent of $\omega$ and real. When non-markovian,   $P(k, \omega)$ develops a non-trivial frequency dependence. When the system is at equilibrium or linear response theory applies, the fluctuation dissipation theorem relates the power spectrum to the imaginary part of a response function guaranteeing that this is a real function \cite{LL}. In cosmology and in non-relaxed situations, the power spectrum can violate these assumptions. This will be discussed in the following section.


\subsubsection{The cosmological spectrum}

In the cosmological case that we consider here, we work in comoving coordinates in a spatially flat FRW universe. In this context, $\vec k$ is just the usual comoving wavevector. We normalise the scale factor to $a(t_0) =1$ today, and $P(\vec k,t)$ should be understood as the matter power spectrum in these comoving coordinates at a given time $t$.

The power spectrum in cosmology is neither Markovian nor time-translation invariant. The full power spectrum involves non-linear effects on short scales below roughly 1 Mpc. Numerical simulations are commonly used to delve into the non-linear regime although semi-analytic models such as the halo model \cite{Asgari:2023mej} can also be very useful. More modestly and only valid on large scales, the linear power spectrum allows one to elaborate on the time-translation breaking effects involved in the growth of structure \cite{Ansari:2024efj,Donath:2023sav}. In a nutshell and at the linear level, structures start growing in the matter dominated era where we neglect their slow logarithmic growth in the radiation era for scales entering the horizon before matter-radiation equality. 
Scales enter the horizon at a time $t_k$ such that
\be 
k= a(t_k) H(t_k)
\ee
where $H$ is the Hubble rate and $a(t)$ the scale factor.
When $t_k\gtrsim t_{\rm eq}$, the density contrast $\delta(\vec k,t)$  grows as
\be 
\delta(\vec k,t) = \frac{D(t)}{D(t_k)} \delta_{\rm ini}(\vec k)
\ee
whilst for $t_k\lesssim t_{\rm eq}$ we have for $t\ge t_{\rm eq}$
\be 
\delta (\vec k,t)= \frac{D(t)}{D(t_{\rm eq})} \delta_{\rm ini}(\vec k).
\ee
As space-translation invariance is preserved, we can use the mixed representation for the power spectrum
\be
\langle j(\vec k, t) j(\vec k',t')\rangle = P(\vec k, t,t') \slashed{\delta}^{(3)}(\vec k+ \vec k')
\ee
where 
\be 
P(\vec k, t,t')= J_0(t) J_0(t') \langle \delta (\vec k,t) \delta (\vec k', t')\rangle
\ee
where $J_0(t)$ is the background density. The initial power spectrum is given by
\be 
\langle \delta_{\rm ini}(\vec k) \delta_{\rm ini}(\vec k')= {\cal P}_{\rm ini}(k) \slashed{\delta}^{(3)}(\vec k+ \vec k')
\ee
and the power spectrum at equal times is
\be 
\langle \delta_{}(\vec k,t) \delta_{}(\vec k',t)= {\cal P}(k,t) \slashed{\delta}^{(3)}(\vec k+ \vec k').
\ee
At unequal times, we can use the growing mode $D$ to relate the power spectrum to the one at equal times. Let us assume that for a given scale we have that $t>t_k$ then when $t'\le t$ we have
\be 
\langle \delta_{}(\vec k,t) \delta_{}(\vec k',t')= \left ( \theta(t'-t_k) \frac{D(t')}{D(t)} + \theta(t_k-t')\frac{D(t_k)}{D(t)} \right ) {\cal P}(k,t) \slashed{\delta}^{(3)}(\vec k+ \vec k').
\label{PP}
\ee
When $t'<t<t_k$, essentially we have 
\be 
\langle \delta_{}(\vec k,t) \delta_{}(\vec k',t')=  {\cal P}_{\rm ini}(k)\slashed{\delta}^{(3)}(\vec k+ \vec k').
\ee
In practice in (\ref{PP}), the equal time power spectrum can be constructed using the transfer function
\be 
{\cal P}(k,t)= T(k,t) {\cal P}_{\rm ini}(k)
\ee
and the linear growing mode can be used to estimate the power spectrum at unequal times. 
The total matter density comprises a background part and the random one so that
\be 
\bar J= J_0 +  j 
\ee
where the background density $J_0$ is only time-dependent i.e. its Fourier transform in space is a Dirac distribution at the origin. 

When a scalar field is coupled to such a random matter distribution, energy could leak from large scales to small scales and lead to dissipation effects. Moreover integrating out the matter distribution on short scales will lead to backreaction effects on the scalar field dynamics.  In practice, the scales that are integrated out go from solar to  galactic. 
Eventually, we will be interested  in the large distance regime for the scalar field evolving on cosmological scales.
This will be studied by defining an  effective field theory for the scalar on cosmological scales obtained after integrating out the scalar response to  the presence of matter on short scales. Typically, we will integrate out scales shorter than $L\sim 1$ Mpc and consider the cosmological dynamics on larger scales $k\le \frac{1}{L}$ up to the horizon scale $H$. When considering the effects of the local dynamics on the background cosmology, we will solve a modified Klein-Gordon equation which becomes space-independent at the background level. At the linear cosmological perturbation level, the scales considered will be up to around $500$ Mpc as relativistic effects at the horizon scale are left for future work. The propagation equation of such perturbations will be affected by the local dynamics in the form of random sources corresponding to the exchange of information between large and small scales, and corrections to the scalar's mass.

\subsection{The "in-in" formalism}
Let us consider the scalar field theory with its self-interactions as a quantum system \cite{Porto:2024cwd, Brax:2024obu}. This is only a matter of convenience here as the dynamics are classical. In essence we will focus on the classical dynamics characterised by the tree level diagrams \footnote{Quantum effects in the context of gauge theories have been recently considered  \cite{Kaplanek:2025moq,Kaplanek:2026kpp} or for stochastic inflation \cite{Cespedes:2026fdp}.}, i.e. the lowest order in the semi-classical loop expansion. A similar approach was recently followed for the propagation of gravitational waves in a stochastic medium \cite{Amoruso:2026txw}. Contrary to the "in-out" formalism, we focus on vacuum expectation values of the scalar field and its time evolution, see \cite{Donath:2024utn} for a modern take on the similarities and differences between the two formulations. When coupling fields to matter, this allows one to take into account possible dissipation effects.
In the interaction picture, where the interaction Hamiltonian is $H_I = \int d^3x V_{\rm eff}(\varphi_I)$, the evolution operator is $U=T\left( \exp(-i \int_{-\infty}^t dt' H_I(t')) \right)$. The effective potential involves a coupling to matter and reads
\be 
V_{\rm eff}(\varphi)= V(\varphi) + \bar J B(\varphi)
\ee
where $\bar J$ is the source term representing the total matter distribution $\bar J= \bar J_0 + j$.  The function $B(\varphi)$ is the coupling function.

The full quantum field $\varphi$ is related to the field in the interaction picture $\varphi_I$ as $\varphi=U^\dag \varphi_I U$, with $U^\dag =\bar{T}\left( \exp(i \int_{-\infty}^t dt' H_I(t')) \right)$. Here $T$ is the T-product that orders the operators with increasing time from right to left and $\bar{T}$ orders the path with decreasing time from right to left too.
Notice that initially the quantum field is taken to be the free field before the interactions are switched on. 
 
\begin{figure}[ht] 
	\includegraphics[scale=0.5]{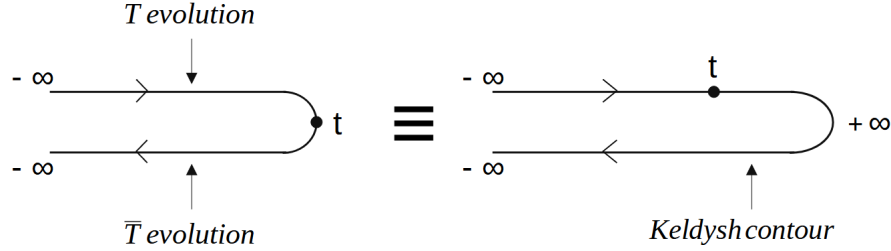}				
	\caption{Keldysh contour with an operator inserted at time $t$ and the contour rejected to infinity thanks to  $U^\dag(t,+\infty)U(t,+\infty)=I$.}
  \label{k}
	\end{figure}

Defining by $\ket{\rm in}$ the vacuum of the free field theory, the expectation values of the quantum field are such that
    \begin{equation}
    \langle {\rm in}\vert \varphi\vert {\rm in}\rangle  =  \langle \rm{in}\vert \bar{T}\left( e^{i \int_{-\infty}^t dt' H_I(t')} \right) \varphi_I T\left( e^{-i \int_{-\infty}^t dt' H_I(t')} \right)\vert \rm{in}\rangle   .  
    \end{equation}
Hence by  reading the evolution of the operators from right to left, we can draw the diagram \ref{k} where the operators in the first $T$ product are on the top line and the ones with the $\bar{T}$ product on the bottom line. We can introduce the generating function of correlators in the "in-in" formalism, $Z[J]=\langle {\rm in}\vert  {\cal T}\left( e^{-i \int d^4x J\varphi} \right)\vert {\rm in} \rangle$ where ${\cal T}$ is the time ordering along the whole Keldysh contour. This can be written as a path integral with  a field evolving with increasing time (to the right, $T$) and another with decreasing time (to the left, $\bar{T}$).	
We assign the subscript 1 to the quantities following the $T$ evolution on the top contour and 2 to the ones following the $\bar{T}$ evolution \cite{Schwinger:1960qe, Keldysh:1964ud} and obtain the generating functional
    \begin{equation}    
    Z[J_1,J_2]=\int \mathcal{D}\varphi_1 \mathcal{D}\varphi_2 e^{iS_1-i\int d^4x\sqrt{-g} J_1\varphi_1 - iS_2+i\int d^4x\sqrt{-g} J_2 \varphi_2})
    \end{equation}
    where $S_{1,2}=S(\varphi_{1,2})$.
    Classically the same  sources $\bar J_{1,2}=\bar J$ are assigned to the top and bottom contours.  We have also $\varphi_1(+\infty)=\varphi_2(+\infty)$ by continuity. 
    
In the case of a scalar tensor theory defined by the coupling function $A(\varphi)$ and normalising the Einstein-frame matter density for $\varphi=\varphi_c$ we have \cite{Brax:2025ahm}
\be 
B(\varphi)= \frac{A(\varphi)}{A(\varphi_c)}-1.
\ee
Typically we will assume that $B$ is small and can be used as an expansion parameter. We will make this expansion scheme explicit below.

As in usual quantum field theory we introduce the generating function of connected diagrams $W$, $Z[J]=e^{iW[J]}$, so here
    \begin{equation}
    Z[J_1,J_2]=e^{iW[J_1,J_2]}
    \end{equation}
    and we perform a Legendre transform to obtain the generating function of one-particle irreducible diagrams
    \begin{equation}
    W[J_1,J_2] = \Gamma[\Phi_1,\Phi_2] - \int d^4x\sqrt{-g} J_1\Phi_1 +\int d^4x\sqrt{-g} J_2\Phi_2,
    \end{equation}
    where the  fields are obtained as
    \begin{equation}
    \Phi_1=-\frac{\delta W}{\delta J_1} \hspace{0,5cm}\text{and}\hspace{0,5cm} \Phi_2=\frac{\delta W}{\delta J_2},
    \end{equation}
    such that the equations of motion for the $\Phi_{1,2}$ fields are
    \begin{equation}
    \frac{\delta\Gamma}{\delta\Phi_1}=J_1 \hspace{0,5cm}\text{and}\hspace{0,5cm} \frac{\delta\Gamma}{\delta\Phi_2}=-J_2 .
    \end{equation}
    When $ J_{1,2}=0 \Rightarrow \Phi_{1,2}=\bar{\Phi}_{1,2}$ which is the classical solution of the quantum field theory.  We denote these solutions by $\bar \Phi_{1,2}$.

    In the following we will modify this setting by introducing two new ingredients. First of all we will integrate out the fields sourced by the local distribution of matter. Then we will average over the matter fluctuations.

\section{The  Wilsonian action} 
\label{sec:wil}
\label{conservative} 
\subsection{Integrating out short distances}    
We will obtain the effective action for the scalar field on cosmological scales in the Schwinger-Keldysh formalism in a two-step approach. First we will integrate  over the field configurations on short distances corresponding to the size of the galactic systems for instance. This will give rise to a Wilsonian action for the long distance fields on cosmological scales for each realisation of the random matter distribution. Then we will take the average over the random matter distribution.  Doing so we assume ergodicity, i.e. for a given realisation of the matter distribution we  consider the values of the large distance field in cells of  size  $L$ over which short distance fluctuations have been integrated out. The averaged value of the large distance field over the different cells in the Universe corresponds to its  statistical average by the ergodic hypothesis. In short, we decompose the field as
\be 
\varphi=\underbrace{\varphi_0}_{\text{short distances}} + \underbrace{\phi}_{\text{large distances}}.
\ee
This is reminiscent of  the usual decomposition between high and low energy field in a Wilsonian treatment of quantum field theory where the short distance physics is eventually integrated out. 
We also separate the random matter distribution as
\be 
\bar J= j_< + j_>
\ee
where $j_<= \bar J_0 + W_< \star j$ has a support on large scales and includes the background matter distribution. The small scale source is $j_>= W_>\star j$. 
For instance, the short distance source is obtained by using a window function $W_>(k)= 1_{k\ge \frac{1}{L}}$ corresponding to distances shorter than the sliding scale $L$ \footnote{ The equations of motions are obtained by decomposing integrals involving the field $\Bar\Phi= W_>\star \Phi$ defined on short scales as  
\be 
\int d^3x dt \bar \Phi (\vec x,t) F(\vec x, t)= \int \slashed{d}^3 k dt \bar \Phi (\vec k,t)  1_{\vert \vec k \vert \ge \frac{1}{L}}F(-\vec k,t)
\ee
implying that the functional derivative with respect to $\Phi$ gives $ 1_{\vert \vec k \vert \ge \frac{1}{L}}F(-\vec k,t)$ in Fourier space or $W_> \star F$ in real space.}
. In real space, we have
\be 
W_<(x)= \int_{\vert \vec k\vert \le \frac{1}{L}} \slashed{d}^3 k e^{i \vec k.\vec x}
\ee
and $W_>(x)= \delta^{(3)}(\vec x) -W_<(x)$. 

When we consider cosmological situations, we will need to match the local geometry around structures to the large scale metric. We will simply assume that local structures are nearly spherical like dark matter halos and locally
\be 
ds^2= a^2(t,r)(-dt^2 + dr^2 + r^2 d\Omega^2)
\ee
with a Lema\^itre-Tolman-Bondi metric where $a(t,r)$ interpolates between $a(t,r)=1$ inside structures where the metric is locally Minkowskian and $a(t,r)=a(t)$ far away where the cosmological setting applies.

Let us start with the generating functional 
\begin{eqnarray}
&& \int \mathcal{D}\varphi_1 \mathcal{D}\varphi_2 e^{iS_1(\varphi_1) - iS_2(\varphi_2) -i\int d^4 x\sqrt{-g}  J_1\varphi_1 +i\int d^4x\sqrt{-g} J_2\varphi_2} \nonumber \\ & 
	& = \int \mathcal{D}\phi_{1} \mathcal{D}\phi_{2} e^{iW(\phi_1,\phi_2,J_1,J_2) -i\int d^4 x\sqrt{-g}  J_1\phi_1 +i\int d^4 x\sqrt{-g} J_2\phi_2}. 
	\nonumber \\
\end{eqnarray}	
 The connected generating functional is given by 
\be 
e^{iW(\phi_1,\phi_2,J_1,J_2)} =\int \mathcal{D}\varphi_{1,0} \mathcal{D}\varphi_{2,0}  e^{iS_1(\varphi_{1,0}+\phi_1) - iS_2(\varphi_{2,0}+\phi_2) -i\int d^4x \sqrt{-g}J_1\varphi_{1,0} + i\int d^4 x \sqrt{-g} J_2\varphi_{2,0}} .
\ee
We can now introduce the effective action $\Gamma$ as the Legendre transform with respect to the large distance fields
    \begin{equation}
    { W}(\phi_1,\phi_2,J_1,J_2) = \Gamma[\phi_1,\phi_2,\Phi_{1,0},\Phi_{2,0}] -\int d^4x\sqrt{-g} J_1\Phi_{1,0} +\int d^4x \sqrt{-g} J_2\Phi_{2,0}
    \end{equation}
where $\Phi_{1,0}=-\frac{\delta {W}}{\delta J_1}, \Phi_{2,0}= \frac{\delta { W}}{\delta J_2} $.
At the classical level when  $J_{1,2}=0$  the classical  fields are given by $\bar{\Phi}_{1,0}=-\frac{\delta W}{\delta J_1}\arrowvert_{J_1=0}$ and $\bar{\Phi}_{2,0}=\frac{\delta W}{\delta J_2}\arrowvert_{ J_2=0}$. They also solve the classical field equations $\frac{\delta \Gamma}{\partial \Phi_1}\vert_{\bar \Phi_{1,2}}=0 $ and  $\frac{\delta \Gamma}{\partial \Phi_2}\vert_{\bar\Phi_{1,2}}=0 $. It is important to notice that the classical solutions $\bar \Phi_{1,2}(\phi_{1,2})$ depend on the fields $\phi_{1,2}$ and the random sources $\bar J_{1,2}$. 

We can now define the grand potential\footnote{By analogy with thermodynamics in open systems where the number of particles is not conserved. } averaged over the matter fluctuations on scales shorter than $L$
\be 
{\rm E}(e^{i\Omega (\phi_1,\phi_2,\xi_{1,2})})= \langle e^{i
\Gamma[\phi_1,\phi_2,\bar \Phi_{1,0},\bar \Phi_{2,0}]}\rangle
\ee
where on the left hand side the grand potential depends on two random fields induced by the openness of the system where information can flow in and out of the local dynamics. The averaging process ${\rm E}(.)$ will be made explicit below. 
The grand potential acts as the effective action which governs the evolution of the scalar on cosmological scales. 

This can be obtained as follows. 
The average value over the random matter distribution can be expanded using the cumulants of $\Gamma$ defined as 
\be 
\ln \langle e^{i
\Gamma[\phi_1,\phi_2,\bar \Phi_{1,0},\bar \Phi_{2,0}]}\rangle
= \sum_{n=0}^\infty i^n\frac{C_n(\Gamma)}{n!}
\ee
where $C_0=1$, $C_1(\Gamma)= \langle \Gamma \rangle$, $C_2(\Gamma)= \langle \Gamma^2\rangle - \langle \Gamma\rangle^2$ and $C_3(\Gamma)= \langle (\Gamma- \langle \Gamma\rangle)^3\rangle$. As the functional $\Gamma$ is a real functional of real fields, the terms of even orders are real and cannot be identified as terms in a real effective action. This signals that the effective description on large distances becomes intrinsically stochastic when integrating over short scales. Let us introduce two classical noises $\xi_{1,2}$ such that
\be 
\ln {\rm E}(e^{i\int d^4x \sqrt{-g} (\xi_2 \phi_2-\xi_1\phi_1) })= \sum_{n=0}^{\infty} (-1)^n \frac{C_{2n}(\Gamma)}{(2n)!}.
\ee
This determines completely all the correlation functions of the two noises by differentiation, i.e. this defines the probability law of the two noises. As a result we obtain the grand potential  on cosmological scales as
\be 
\Omega (\phi_1,\phi_2,\xi_{1,2})=\sum_{n=0}^\infty  (-1)^n \frac{C_{2n+1}(\Gamma)}{(2n+1)!} +\int d^4x \sqrt{-g} (\xi_2 \phi_2-\xi_1\phi_1).
\ee
This grand potential is not local and involves multiple integrals which could  lead to tail effects  for instance .
One can always decompose the grand potential as 
\be 
\Omega (\phi_1,\phi_2) = {\cal S}(\phi_1)-\int d^4x\sqrt{-g}  \xi_1 \phi_1 - {\cal S}(\phi_2)+\int d^4x \sqrt{-g}\xi_2 \phi_2 + {\cal R} (\phi_1,\phi_2)
\ee
where we have separated a term which is a difference of two non-local  actions depending on the single functional ${\cal S}$ and a remaining term involving the couplings between $\phi_1$ and $\phi_2$.
The classical scalar dynamics are obtained after defining 
\be 
\phi_+=\frac{\phi_1+\phi_2}{2}, \ \phi_-=\phi_1 -\phi_2
\ee
such that
\be 
\Omega (\phi_+,\phi_-)= {\cal S}(\phi_++ \frac{\phi_-}{2})- {\cal S}(\phi_+- \frac{\phi_-}{2})+R(\phi_++ \frac{\phi_-}{2},
\phi_+- \frac{\phi_-}{2})- \int d^4x \sqrt{-g} (\xi_+ \phi_-+ \xi_- \phi_+) .
\ee
Imposing the classical equations of motions corresponds to the Klein-Gordon equation for the field $\phi_+$  with $\phi_-=0$. On cosmological scales $\frac{\delta \Omega}{\delta \phi_-}\vert_{\phi_-=0}=0$
implies the Klein-Gordon equation for the large distance field
\be 
\frac{\delta {\cal S}(\phi)}{\delta \phi}=\xi_++ \frac{1}{2}( \frac{\delta R(\phi_1,\phi_2)}{\delta \phi_2}- \frac{\delta R(\phi_1,\phi_2)}{\delta \phi_1})\vert_{\phi_1=\phi_2=\phi}.
\label{dyna}
\ee
The terms involved in $R$ represent radiation-reaction effects whilst the noise source corresponds to the openness of the system from large to small scales and vice-versa. This equation is in general non-local and contains memory effects. The dynamics of the light scalar fields on large scales are determined by (\ref{dyna}).

\subsection{The short distance field}

We will now focus on determining the dynamics of  light scalars coupled to matter and carry out the procedure outlined in the previous section, i.e. we will obtain the one-particle irreducible functional $\Gamma$ and analyse the characteristics of the two noises $\xi_{1,2}$.
The first step in obtaining the grand potential for the cosmological field is to solve for the local dynamics. 
The Klein-Gordon equation for the classical fields on short distances is given by
    \begin{equation}
    \begin{aligned}
     \begin{array}{l}
	\square \bar{\Phi} = W_>\star( j_> B'(\phi+ \bar\Phi)+ V'( \phi+ \bar\Phi)).
\end{array}.
    \end{aligned}
    \end{equation}
The local field is only sourced by the local density $j_>$ and not the cosmological one $\bar J_0$. 
As we are interested in classical configurations, we have $\Bar\Phi_1=\bar\Phi_2=\bar \Phi$. Moreover, the short distance field depends on the classical cosmological field $\phi_1=\phi_2=\phi$. They also coincide with $\bar \Phi_+$ and $\phi_+$.
The Klein-Gordon equation is a non-linear equation which is very hard to solve in general so we will consider two simplified situations with cosmological applications.
\subsubsection{The unscreened case}
When no screening takes place, i.e. the local structures act only as perturbative corrections to the large scale dynamics,  the cosmological field penetrates inside the local structures and serves as a background field everywhere in the Universe. This is a situation where the influence of the local matter density on the dynamics of the scalar field is mild and can be treated as a local perturbation.

We can then expand the coupling function as
\be 
B(\phi+ \Phi)=  \sum_{n\ge 1} \frac{B_n(\phi)}{n!} (\frac{\beta}{m_{\rm Pl}})^{n} \Phi_{}^n
\ee
and the interaction potential
\be 
V(\phi+ \Phi)= \sum_{n\ge 0}\frac{V_n(\phi)}{n!}(\frac{\beta}{m_{\rm Pl}})^{n} \Phi^n
\ee
where we have assumed that the two potentials are functions of $\beta(\phi+\Phi)/m_{\rm Pl}$ corresponding to a single suppression scale given by $m_{\rm Pl}/\beta$ for higher order terms in the expansion. This applies for instance to potentials in $e^{-4\beta \phi/m_{\rm Pl}}$ and coupling functions in $e^{-\beta \phi/m_{\rm Pl}}$ for dilatons \cite{Smith:2025grk}. 
This leads to 
\begin{equation}
    \begin{aligned}
     \begin{array}{l}
	(\square-m^2(\phi))\bar{\Phi} = W_{>}\star({\cal J}(\phi) +\sum_{n\ge 2} (\frac{\beta}{m_{\rm Pl}})^n \frac{D_{n+1}(\phi)}{n!}\Phi^n ) 
\end{array}.
    \end{aligned}
    \end{equation}
where the source is defined by ${\cal J}(\phi)=\frac{\beta}{m_{\rm Pl}}(j_> + V_1(\phi))$ and higher order corrections read
$D_n(\phi_1)= B_n(\phi) j_> +V_n(\phi)$. The mass of the short distance field is given by
\be 
m^2(\phi)= (\frac{\beta}{m_{\rm Pl}})^2 (V_2(\phi) + j_> B_2(\phi)).
\label{mass}
\ee
In the following, we will consider that the time and spatial vatiations of the large scale cosmological fields are small compared to the short distance scales which are integrated out. As a result, we can consider $\phi$ to be nearly constant in the previous equations. 
The Klein-Gordon equation can then be solved by iteration
\be 
\bar \Phi =\sum_{n\ge 0}\bar\Phi ^{(n)}
\ee
where the first step is simply
\begin{equation}
    \begin{aligned}
     \begin{array}{l}
	(\square-m^2(\phi))\bar{\Phi}^{(0)} = W_{>}\star {\cal J}(\phi) 
\end{array}
    \end{aligned}
    \end{equation}
and the following steps are such that
\begin{equation}
    \begin{aligned}
     \begin{array}{l}
	(\square-m^2(\phi))\bar{\Phi}^{(n+1)} = W_{>}\star ( \sum_{p\ge 2} (\frac{\beta}{m_{\rm Pl}})^p \frac{D_{p+1}}{p!}(\bar\Phi^{(n)})^p)
\end{array}.
    \end{aligned}
    \end{equation}
At each step the Green's function corresponds to the radiation from a classical source, see appendix \ref{app:green}, 
\begin{equation}
    \bar{\Phi}^{(0)}(x) = \int d^4x'  G_{\rm ret}(x-x')(W_{>}\star{\cal J})(x') .
\end{equation}
The retarded propagator is simply
\be 
G_{\rm ret}(\vec x,t)= \int \slashed{d} \omega \slashed{d}^3 p e^{-i\omega t+i\vec p.\vec x} \frac{1}{(\omega +i\epsilon)^2 -\vec p^2 - m^2(\phi)}.
\ee
Here we assume that locally the Universe is Minkowskian in the local structures and the Green's functions can be taken to be the flat ones. 

Let us distinguish two cases. The first one corresponds to time-translation invariant processes for which 
we can use
\be 
{\cal J}(\vec p,\omega)=\frac{\beta}{m_{\rm Pl}}(j_> (\vec p,\omega ) + V_1(\phi) \slashed{\delta}(\omega)\slashed{\delta}^{(3)}(p))
\ee
where the last term comes from the background taken to be nearly constant on the scales considered here.
The momentum integrals are taken to be for $\ \vert \vec p \vert \ge \frac{1}{L}$ where $L$ is the distance defining the short wavelength modes. As a result  
we have 
\be 
\bar{\Phi}^{(0)}(\vec x,t)= -\frac{i\beta}{m_{\rm Pl}}  \int_{\vert \vec p\vert \ge \frac{1}{L}} \slashed{d}^3 p  \frac{1}{2 \omega (p)} (j(\vec p,\omega (p)) e^{i\vec p.\vec x-i\omega(p) t}- j(\vec p, -\omega (p))e^{i\vec p.\vec x+i\omega(p) t})
\ee
where $\omega(p)= \sqrt{\vec p^2 +m^2(\phi)}$. The two contributions come from the two poles of the retarded propagator. We can write the Fourier transform in a compact form as
\be 
\bar \Phi^{(0)}(\vec p,\omega)= -\frac{i\beta}{m_{\rm Pl}}
\frac{j(\vec p,\omega)}{2 \omega}\slashed{\delta}(\vert \omega\vert - \omega(p))1_{\vert\vec p\vert \ge \frac{1}{L}}
\ee
where the $\delta$-function selects the two choices $\pm \omega(p)$ leading to the sign difference between the two exponentials. 
As a result, the short distance field becomes a Gaussian random process. Higher order terms are obtained by iteration using the retarded  propagators. 
The solution can be ordered as a power series in $\beta/m_{\rm Pl}$. The leading order term $\bar \Phi^{(0)}$ is of order $\beta/m_{\rm Pl}$ and contains terms of order $(\beta/m_{\rm Pl})^{n+1}$ when expanding the propagators in powers of $m^2(\phi)$. The next term $\Phi^{(1)}$ starts at order $(\beta/m_{\rm Pl})^{4}$. The power counting can be easily performed and at a given order there are only a finite number of terms. 

The second case is the cosmological setting where time-translation is broken. It is convenient to work in a mixed representation where the Fourier transform only applies to space and not time. Then we  have
\be 
G_{\rm ret}(\vec k,t)= \theta(t) \int \slashed{d}^3 p \frac{1}{\omega (p)} \sin (\omega (p) t ) e^{i\vec p.\vec x}
\ee
where we have integrated over $\omega$ in the lower half-plane where convergence at infinity is guaranteed when $t>0$ and two poles lie below the real axis. When $t<0$, there are no poles in the upper-half plane and the Green's function vanishes. This leads to 
\be 
\bar\Phi^{(0)}(\vec x ,t)= \frac{\beta}{m_{\rm Pl}}\int_{\vert \vec p\vert \ge \frac{1}{L}} \slashed{d}^3 p \frac{1}{\omega (p)}\int_{-\infty }^t \sin (\omega(p)(t-t')) j(\vec p,t') dt'.
\label{scalba}
\ee
This corresponds to the Fourier transform
\be 
\bar\Phi^{(0)}(\vec p,t)= \frac{\beta}{m_{\rm Pl}}\frac{1_{\vert\vec p\vert \ge \frac{1}{L}}}{\omega (p)}\int_{-\infty }^t \sin (\omega(p)(t-t')) j(\vec p,t') dt'.
\ee
This will be useful to characterise the corrections to the cosmological background equations. 

\subsubsection{Screening}
When screening takes place locally, the short distance field is determined by the local matter density. Screening takes places when the Compton wavelength of the scalar field is much shorter than the size of the local matter over-densities. The field thus settles at the minimum of the effective potential and satisfies
\be j_> B'(\bar\Phi)+ V'(\bar\Phi )=0.
\ee
This determines the local field as a function of the local matter density. This  takes place only  in the screened regions of space. In other parts of the Universe, the unscreened case applies. We will consider below that all the local structures are screened as this simplifies the mathematical treatment and exemplifies the drastic effects of screening on the long distance field. The case where only dense enough regions lead to screening is more involved and is left for future work.

\section{The grand potential and the noise}
\label{sec:grand}
\subsection{The grand potential}

As we are considering a classical system, the one-particle irreducible functional is simply the classical action where the short distance field $\Phi$ is taken to be its classical value solving the short distance equations of motion
\be 
\Gamma[\phi_1,\phi_2,\bar \Phi,\bar \Phi]= S(\phi_1+\bar \Phi)-S(\phi_2+\bar \Phi)
\ee
where classically we have $\bar \Phi_{1,2}= \bar \Phi$. This is the lowest order in the semi-classical expansion of the path integral and corresponds to the tree level diagrams with no loops involved. We use the fact that fundamentally we are dealing with a classical problem incorporating  the random fluctuations of the source $j$. The path integral formalism is useful to extract features such as the noise corresponding to the exchange of information between short and large scales, or the interaction between the short scale physics and the large scales leading to new effective interactions  after integrating out the short range dynamics.

It is convenient to use the $\pm$ basis for the fields. Recall that the short distance field is a classical field $\bar\Phi_+\equiv \bar\Phi$ and is a function of the classical cosmological field $\phi_+$. 
As a results we have 
\be 
S(\phi_++\bar\Phi_+ \pm \frac{\phi_-}{2})= \sum_{n\ge 0}\frac{(\pm)^n}{2^n n!} \int d^4x_1\dots d^4x_n \frac{\delta^n S}{\delta\phi(x_1)\dots  \delta \phi(x_n)}\vert_{\phi_++\bar\Phi_+}\phi_-(x_1)\dots \phi_-(x_n)
\ee
involving the functional derivatives of the classical action $S$. 
The one-particle irreducible functional is then 
\be 
\Gamma[\phi_1,\phi_2,\bar \Phi_{1,0},\bar \Phi_{2,0}]=\sum_{n\ge 0}\frac{1}{2^{2n} (2n+1)!} \int d^4x_1\dots d^4x_{2n+1}\frac{\delta ^{2n+1} S}{\delta \phi(x_1) \dots \delta \phi(x_{2n+1})}\vert_{\phi_++\bar\Phi}\phi_-(x_1) \dots \phi_-(x_{2n+1}).
\ee
This is a sum of odd powers of $\phi_-$ as the even powers cancel out.

All the cumulants of odd order $C_{2p+1}(\Gamma)$ of order greater than three corresponding to $p\ge 1$ can be expanded in powers of $\phi_-$ with no linear term in $\phi_-$.  This is crucial as
the cosmological equations of motion are obtained by taking the first functional derivative of these cumulants  with respect to $\phi_-$ and setting $\phi_-=0$. We see immediately that the only term which leads to a non-vanishing value is $C_1(\Gamma)=\langle \Gamma\rangle.$ In conclusion, we find that the only terms with a non-vanishing contribution of the grand potential to the classical equations of motion  are
\be 
\Omega (\phi_1,\phi_2,\xi_{1,2})=\langle \Gamma \rangle(\phi_1,\phi_2) +\int d^4x \sqrt{-g}(\xi_2 \phi_2-\xi_1\phi_1).
\ee
This characterises all the features of the  dynamics on large scales and contains two terms. The first term accounts for the non-trivial interactions between large scale fields induced by short scales. This is the effective action. The second is a feature of the openness of the system and corresponds to a Langevin source.

Moreover we have
\be 
\langle \Gamma \rangle(\phi_1,\phi_2) =\langle S(\phi_1+\bar \Phi)\rangle -\langle S(\phi_2+\Phi)\rangle
\ee
implying that there is no reaction-radiation term (\ref{dyna})
\be 
R(\phi_1,\phi_2)=0
\ee
and
\be 
{\cal S}(\phi)= \langle S(\phi+\Bar \Phi) \rangle.
\ee
We conclude that the grand potential on cosmological scales is obtained by solving for the local field $\bar \Phi$ and averaging the action $S$. There is also a noise term which characterises the openness of the system.

\subsection{Some properties of the noise}

We can rewrite the generating functional of correlations for the two noises $\xi_\pm$ as
\be 
\ln {\rm E}(e^{-i\int d^4x \sqrt{-g}(\xi_+ \phi_-+\xi_-\phi_+) })= \sum_{n=0}^{\infty} \frac{(-1)^n}{(2n)!} {C_{2n}(\Gamma)}
\ee
where all the cumulants are functions of $\phi_+\pm \frac{\phi_-}{2}$. The correlation functions of the two noises can be obtained by differentiating the generating functional. In fact, we are interested in the correlation function of the physical noise $\xi_+$ obtained by differentiating with respect to $\phi_-$. This can be obtained by Taylor expanding 
\begin{eqnarray}
&&\sum_{n=0}^{\infty} \frac{(-1)^n}{(2n)!} {C_{2n}(\Gamma)}(\phi_++ \frac{\phi_-}{2}, \phi_+- \frac{\phi_-}{2})= \nonumber \\ &&\sum_p \frac{(-i)^p}{p!}\int (\prod_{i=1}^p d^4x_i \sqrt{-g_i} )E(\xi_+(x_1)\dots \xi_+(x_p))\phi_-(x_1)\dots \phi_-(x_p).\nonumber \\
\end{eqnarray}
As the noise is real we expect that all the odd order correlation functions vanish. Indeed, the cumulants of even powers involve even powers of $\Gamma$ and therefore are an even function of $\phi_-$. As a result, all the correlation functions of the noise of odd order vanish. This is in particular the case of the average
\be 
E(\xi)=0.
\ee
This implies that on average the stochastic source of the cosmological equations on large scales vanishes. In the following, the stochastic noise will be treated as a perturbation with no effect on the cosmological background evolution.

\subsection{Evaluating the effective action $\langle \Gamma \rangle$}

It is convenient to use the $\pm$ basis for the fields and rewrite the one-particle irreducible functional as
\begin{eqnarray}
&&\Gamma (\phi_+,\phi_-, \bar\Phi)= -\int d^4x \sqrt{-g} ( \partial \phi_-\partial\phi_+ + \nonumber \\ && V(\phi_+ +\bar \Phi+ \frac{\phi_-}{2})-V(\phi_+ +\bar \Phi- \frac{\phi_-}{2})+ \bar J (B(\phi_++ \bar \Phi+ \frac{\phi_-}{2})-B(\phi_++ \bar \Phi- \frac{\phi_-}{2})))\nonumber \\
\label{effect}
\end{eqnarray}
where the terms in $\bar \Phi$ vanish as $\bar \Phi_-=0$ and there is no kinetic coupling between $\Phi$ and $\phi$ as they have disjoint supports in Fourier space. We must distinguish between the unscreened and screened cases.

\subsubsection{Unscreened short scales}
In the unscreened case, we can expand around the cosmological field and get
\be 
\Gamma= \sum_{n\ge 0} \Gamma^{(n)}
\ee
in powers of $\bar \Phi$. We have for the leading term
\begin{eqnarray}
&&\Gamma^{(0)} (\phi_+,\phi_-, \bar\Phi)= -\int d^4x \sqrt{-g} ( \partial \phi_-\partial\phi_+ + \nonumber \\ && V(\phi_+ +\frac{\phi_-}{2})-V(\phi_+ - \frac{\phi_-}{2})+ \bar J (B(\phi_++ \frac{\phi_-}{2})-B(\phi_+- \frac{\phi_-}{2})))\nonumber \\
\end{eqnarray}
and the corrections are given by
\be 
\Gamma^{(n)} (\phi_+,\phi_-, \bar\Phi)=- \int d^4 x\sqrt{-g}  (\frac{\beta}{m_{\rm Pl}})^n\frac{\bar \Phi^n}{n!}( V_n(\phi_+ +\frac{\phi_-}{2})-V_n(\phi_+ - \frac{\phi_-}{2})+ \bar J (B_n(\phi_++ \frac{\phi_-}{2})-B_n(\phi_+- \frac{\phi_-}{2}))).
\ee
Taking the average and the lowest order terms in $\beta$ we have
\be 
\langle \Gamma\rangle \sim \langle \Gamma^{(0)}\rangle - \frac{\beta}{m_{\rm Pl}}\int d^4x \sqrt{-g} \langle \Bar \Phi \bar J\rangle ( B_1(\phi_++ \frac{\phi_-}{2})-B_1(\phi_+- \frac{\phi_-}{2}))
\label{potL}
\ee
where $\langle \Gamma^{(0)}\rangle $ only depends on $j_<$ as $\langle j_>\rangle =0$ and $\langle \bar \Phi\rangle =0$.
As a result, the Klein-Gordon equation on large scales becomes
\be 
\Box \phi= W_<\star (\partial_\phi (V(\phi) + j_< B(\phi))+ (\frac{\beta}{m_{\rm Pl}})^2 \langle \bar J \bar \Phi\rangle {\cal B}_2(\phi)+\xi)
\label{KGU}
\ee
where
$\frac{dB_1( \phi)}{d\phi}= \frac{\beta}{m_{\rm Pl}} {\cal B}_2 ( \phi)$.
We have taken the long distance limit by convolving with $W_<$. The cosmological evolution on large scale is affected by the dynamics on short scales in two ways. First, there is the noise contribution coming from the exchange of information between short and large scales. There is also a new potential term coming from the correlation between the random source $\bar J$ and the classical solutions $\bar \Phi$. This new potential is suppressed by two powers of $\beta$ and should be a small correction. On the other hand, even a small correction to the potential could modify the asymptotic dynamics of $\phi$ where it has both small kinetic and potential energies, i.e. being a test field.  We will take them into account in turns and evaluate the scale-dependent coefficient
\be 
A(L)\equiv  W_<\star \langle \bar J \bar \Phi\rangle (\vec x, t) 
\ee
obtained by averaging over short distance fluctuations.

\subsubsection{The correction to the Klein-Gordon equation: unscreened case}
Let us evaluate the coefficient of the leading correction to the long distance equation in the time-translation invariant case 
where
\be 
A(L) = \int_{\vert \vec p \vert \le \frac{1}{L}} \slashed{d} \omega \slashed{d}^3 p  e^{i\vec p.\vec x-i\omega t} \langle \int \slashed{d} \omega' \slashed{d}^3 k \bar J(\vec k, \omega') \bar \Phi (\vec p-\vec k, \omega-\omega')\rangle .
\ee
Now using that  space-time invariance is respected, we have 
\be 
\langle j(\vec k,\omega ) j(\vec p, \omega')\rangle = P(\vec k,\omega ) \slashed{\delta}^{(3)}(\vec k+ \vec p)\slashed{\delta}
(\omega+\omega')
\ee
and
\be 
\Bar \Phi (\vec p,\omega)= -i\frac{\beta}{m_{\rm Pl}}\frac{j(\vec p, \omega)}{2\omega} \slashed{\delta}(\vert \omega\vert  -\omega(p))1_{\vert \vec p\vert \ge \frac{1}{L}}
\ee
where $\omega(p)= \sqrt{\vec p^2 +m^2(\phi)}$. This implies that 
\be 
A(L)=-\frac{i\beta }{m_{\rm Pl}} \int_{\vert \vec p \vert \ge \frac{1}{L}} {\slashed d}^3 p \frac{P(\vec p, \omega (p))-P(\vec p, -\omega (p))}{2\omega (p)}
\ee
which depends on $L$. Notice that as $P$ is a function of $\vert  \vec p\vert$ to preserve isotropy, this is also
\be 
A(L)= \frac{\beta }{m_{\rm Pl}} \int_{\vert \vec p \vert \ge \frac{1}{L}} {\slashed d}^3 p \frac{\Im (P(\vert \vec p\vert , \omega (p)))}{\omega (p)}
\ee
where $\omega(p)$ depends only on the non-fluctuating part of $m(\phi)$ at leading order, see (\ref{mass}). 
In the Markovian case and for equilibria satisfying the fluctuation-dissipation theorem, this reduces to 
\be 
A(L)= 0.
\ee
The correction to the equations of motion only appear for non-Markovian processes. 

In the cosmological case, it is more convenient to use a mixed representation and write
\be 
A(L)= \int_{\vert \vec p \vert \le \frac{1}{L}}  \slashed{d}^3 p  e^{i\vec p.\vec x} \langle \int \slashed{d}^3 k \bar J(\vec k, t) \bar \Phi (\vec p-\vec k, t)\rangle .
\ee
This can written as 
\be 
A(L)=\frac{\beta}{m_{\rm Pl}} J_0(t)\int_{\vert \vec k \vert \ge \frac{1}{L}}\frac{\slashed{d}^3 k }{\omega(k)}
\int_{-\infty}^t \sin (\omega(k) (t-t')) J_0(t') {\cal P}(\vec k,t,t') dt'.
\label{AA}
\ee
This integral involves short distances and therefore the non-linear part of the power spectrum. This could be calculated numerically for instance. Here we will give what happens when we extrapolate the linear power spectrum to non-linear scales as an illustration.

We can decompose the $k$ integral by separating the cases $t_k \ge t$ from $t_k\le t$, i.e. $A(L)=I_1+I_2$. In the first case we have
\be 
I_1=\frac{\beta}{m_{\rm Pl}} J_0(t)\int_{(\vert \vec k \vert \ge \frac{1}{L})\cap (t\le t_k)}\frac{\slashed{d}^3 k }{\omega(k)}{\cal P}_{\rm ini}(\vec k)
\int_{-\infty}^t \sin (\omega(k) (t-t')) J_0(t')  dt'
\ee
and in the second case
\be 
I_2= \frac{\beta}{m_{\rm Pl}}J_0(t)\int_{(\vert \vec k \vert \ge \frac{1}{L})\cap (t\ge t_k)}\frac{\slashed{d}^3 k }{\omega(k)} {\cal P}_{\rm ini}(\vec k) f(t,k)
\ee
where
\be 
f(t,k)
= \int_{-\infty}^{T_k} \sin (\omega(k) (t-t')) J_0(t') + \int_{T_k}^t \sin (\omega(k) (t-t')) \frac{D(t')}{D(t)}T(k,t)
\ee
where $T_k= \max (t_{\rm eq},t_k)$ as there is no significant growth in the radiation era. As can be seen, this is non-vanishing and time-dependent.

In these two cases, the background cosmological equations are therefore modified and become
\be 
\ddot {\bar \phi} + 3 H \dot {\bar \phi}+ \partial_\phi (V(\phi) + J_0 B(\phi))\vert_{\phi=\bar\phi} + A(L) (\frac{\beta}{m_{\rm Pl}})^2 {\cal B}_2 (\bar \phi)=0
\ee
where we have taken into account that on average the noise term  is not present. The noise $\xi$ can be considered as a stochastic perturbation. It will appear in the perturbative equations. 
\subsubsection{Corrections to the potential}
\label{sec:bac}
The main effect of the short distance fluctuations on the large scale dynamics are encapsulated in  the correction to the scalar potential (\ref{potL})
\be 
\delta V= \frac{\beta}{m_{\rm Pl}} A(L) B_1(\bar\phi).
\ee
When the model has a quadratic coupling to matter, the coefficient $B_1(\bar \phi)$ is linear
\be 
B_1(\bar \phi)= 1+ \frac{\beta}{m_{\rm Pl}} {\cal B}_2 \bar\phi.
\ee
A particularly interesting situation corresponds to the case when the light scalar field has no self-interactions
\be
V_{\rm eff}(\bar \phi)= \frac{1}{2} m_0^2 \bar\phi^2 + \rho_0 \frac{\beta}{m_{\rm Pl}}( \bar \phi + \frac{\beta}{2m_{\rm Pl}} {\cal B}_2 \bar\phi^2)
\ee
where $\rho_0$ is the cosmological matter density. This can be written as
\be 
V_{\rm eff}(\bar \phi)= \frac{1}{2}m_\phi^2 (\bar \phi -\hat \phi) -\frac{1}{2}m^2_\phi \hat \phi^2
\ee
where $ \hat \phi= - \frac{\beta \rho_0}{m_{\rm Pl}m^2_\phi}$ and $m^2_\phi= m_0^2 + \frac{\beta^2}{m^2_{\rm Pl}} {\cal B}_2 \rho_0$ is the mass squared of the cosmological field. We assume that $m_0$ is large enough compared to the Hubble rate. In this case, the scalar field tracks the minimum of the effective potential. 
The correction to the effective potential from short distance fluctuations   induces a small change of the  minimum
\be 
\delta \hat \phi= -\frac{\beta^2}{m^2_{\rm Pl}m^2_\phi}{\cal B}_2 A(L)
\ee
and a change in the vacuum energy at order $\beta^2$
\be 
V_0 = -\frac{1}{2}m^2_\phi \hat \phi^2 +\frac{\beta}{m_{\rm Pl}} A(L).
\label{cosmo}
\ee
The short distance fluctuations induce a matter-dependent cosmological constant. This contribution becomes significant when structures form, i.e. at late times. This could provide a solution to the coincidence problem explaining why the measured cosmological constant becomes relevant only at small redshift. 
Of course, this can be interesting only if the induced cosmological constant is of the correct order of magnitude, i.e. close to the matter density of the Universe.
Dimensionally, we can estimate (\ref{AA})
\be 
A(L) \sim \frac{\beta}{m_{\rm Pl}} \rho^2 L^2
\ee
where $\rho$ is the local matter density and $L$ the typical size of the structures. This is valid as long as $\omega (p) \sim p$ corresponding to $m_\phi L \ll 1$ and the factor $1/\omega (p)$ makes the integral sensitive to its lower bound at $p\sim L^{-1}$. In other words,  we have used $\Box^{-1} \sim L^2$. The induced vacuum energy density due the short distance fluctuation is of order then
\be 
\delta V_0 = \frac{\beta}{m_{\rm Pl}} A(L) \sim \frac{\beta^2}{m_{\rm Pl}^2} \rho^2 L^2.
\ee
Typical structures of scales $L\sim 10^{-4} H_0^{-1}$ corresponding to Mpc scales, and local densities $\rho \sim 10^6 \rho_{\rm now}$ like in the Milky Way lead to
\be 
\delta V_0 \sim 10^{4} \beta^2 \frac{\rho_{\rm now}}{m_{\rm Pl}^2 H_0^2} \rho_{\rm now}
\ee
where the Friedmann equation gives $\rho_{\rm now} \sim H_0^2 m_{\rm Pl}^2$ for a matter fraction of around one quarter. For matter couplings $\beta\sim 10^{-2}$ saturating the bound on the coupling between dark matter and scalars \cite{Archidiacono:2022iuu}, we find $\delta V_0 \sim \rho_{\rm now}$. As a result, this mechanism gives a cosmological constant whose value is comparable to the present matter density. Moreover the negative contribution in (\ref{cosmo}) implies that the vacuum energy will have a tendency to increase at small redshift when the cosmological matter density decreases. This would lead to an equation of state smaller than minus one at small redshift and therefore a crossing of the phantom divide. A more thorough investigation is certainly necessary to analyse the validity of this scenario \cite{Contaldi:2026vsw} and corroborate the estimates of the factor $A(L)$.

\subsubsection{Screened short scales}

In the screened case, the large scale fields are perturbations of the short distance $\bar \Phi$ and we expand the effective action using 
\be 
B(\phi+ \Phi)=  \sum_{n\ge 1} \frac{B_n(\Phi)}{n!} (\frac{\beta}{m_{\rm Pl}})^{n} \phi^n
\ee
and the interaction potential
\be 
V(\phi+ \Phi)= \sum_{n\ge 0}\frac{V_n(\Phi)}{n!}(\frac{\beta}{m_{\rm Pl}})^{n} \phi^n.
\ee
We assume that the expansion is valid as long as $\beta \frac{\phi}{m_{\rm Pl}}\ll 1$ where $\beta$ can be large, i.e. this large coupling between scalar and matter is the raison d'\^etre of screening and should be screened to avoid large effects of modified gravity. 
Using (\ref{effect}) and keeping the action to quadratic order for simplicity we find
\be 
\Gamma (\phi_+,\phi_-, \bar\Phi)\simeq -\int d^4x \sqrt{-g} ( \partial \phi_-\partial\phi_++ \frac{\beta}{m_{\rm Pl}}  B_1(\bar\Phi)  j_< \phi_-  +(\frac{\beta}{m_{\rm Pl}})^2( V_2(\bar \Phi)+ \bar J B_2(\bar \Phi) )\phi_- \phi_+ )
\ee
implying that the cosmological dynamics are governed by the Klein-Gordon equation
\be 
\Box \phi=W_<\star (\frac{\beta}{m_{\rm Pl}} \langle B_1(\bar\Phi)\rangle  j_< + (\frac{\beta}{m_{\rm Pl}})^2 (\langle V_2(\bar \Phi)+ \bar J B_2(\bar \Phi) \rangle \phi) +\xi ).
\label{KGSS}
\ee
On cosmological scales where the noise acts as a perturbation, the background evolves according to
\be 
\ddot {\bar\phi} +3 H \dot {\bar \phi}+ M^2 \bar\phi=- \frac{\beta}{m_{\rm Pl}} \bar J_0 \langle B_1(\bar\Phi)\rangle
\label{KGS}
\ee
where the effective mass is given by
\be 
M^2=  (\frac{\beta}{m_{\rm Pl}})^2 \bar J_0 \langle B_2 (\bar \Phi)\rangle + (\frac{\beta}{m_{\rm Pl}})^2 \langle V_2(\bar \Phi)+ j_> B_2(\bar \Phi) \rangle.
\ee
The averages are over the short distance fluctuations.

Screening requires that the mass $M$ is much larger than the Hubble rate \cite{Brax:2021wcv}, as a result the field profile is so tied to the distribution of matter that its dynamics on large scales cannot evolve very efficiently.
The field becomes a simple massive field on large scales. Some dynamics remain coming from the source term in (\ref{KGS}) leading to a drift on large scales
\be 
\frac{\beta \bar\phi}{m_{\rm Pl}}\sim - \frac{\beta^2}{m_{\rm Pl}^2 M^2} \bar J_0 \langle B_1(\bar\Phi)\rangle ={\cal}{O}(\frac{\beta^2 H_0^2}{M^2})
\ee
which is very small as $M\gg H_0$ for screening and thus justifies that the scalar field is essentially determined by the matter distribution and equal to $\bar\Phi$.

\section{Large Scale Perturbations}
\label{sec:pert}

\subsection{Scalar perturbations}
 We focus on perturbations in a fixed Minkowski background in order to capture the physics of the models locally. This should also correspond to the behaviour of perturbations in the quasi-static approximation and on sub-horizon scales \cite{Brax:2004qh}.  Coupling to metric perturbations in an expanding Universe is left for future work. We will distinguish the screened and unscreened cases below. 
\subsubsection{Unscreened}
The perturbed Klein-Gordon equation (\ref{KGU}) reads
\be 
\Box \delta \phi= W_<\star ( (m^2_0 + (\frac{\beta}{m_{\rm Pl}})^3 A(L) B_3(\bar \phi)) \delta \phi +\frac{\beta}{m_{\rm Pl}} B_1 (\bar \phi) \delta j_<+ \xi)
\ee
where $\phi =\bar\phi +\delta\phi $ and $\delta j_<= j_<-\bar J_0$. The mass term is 
\be 
m_0^2= \partial_\phi^2 (V(\phi) + \bar J_0 B(\phi))\vert_{\phi=\bar\phi}.
\ee
The full cosmological mass is corrected by the interaction with the short distance matter distribution
\be 
M^2= m^2_0 + (\frac{\beta}{m_{\rm Pl}})^3 A(L) B_3(\bar \phi).
\ee
The source term comprises two terms. The first is the forcing by the matter fluctuations on large scales and the second by the noise.
In Fourier space, the perturbation equation can be solved and yields
\be 
\delta \phi= \frac{\frac{\beta}{m_{\rm Pl}} B_1(\bar \phi) \delta j_< (\vec k,\omega) +\xi(\vec k,\omega )}{\omega^2-\vec k^2 -M^2} 1_{k\le \frac{1}{L}}
\label{pert1}
\ee
where the cosmological mass term $m_0$ is expected to be of the order of the Hubble rate $H$ for a light scalar field. For quasi-static perturbations with $\omega \sim 0$ and deep in the horizon with $\vert \vec k\vert \gg H$, the fluctuations  are not dampened as $M\sim H$. 
\subsubsection{Screened}
In the screened case and from (\ref{KGSS}) we have directly 
in Fourier space 
\be 
\delta \phi= \frac{\frac{\beta}{m_{\rm Pl}} \langle B_1(\bar \Phi)\rangle  \delta j_< (\vec k,\omega) +\xi(\vec k,\omega)}{\omega^2-\vec k^2 -M^2} 1_{k\le \frac{1}{L}}.
\label{pert2}
\ee
We now  find that if $M\gg H$ the quasi-static perturbations are strongly Yukawa suppressed by the large mass as
\be 
\delta \phi (\vec x,t )\sim -\frac{1}{4\pi}\int_{\vert \vec x- \vec y\vert \le L} d^3 y \frac{e^{-M\vert \vec x-\vec y\vert}}{\vert \vec x- \vec y\vert}(\frac{\beta}{m_{\rm Pl}} B_1(\bar \Phi) \delta j_< (\vec y,t)+\xi(\vec y,t))
\ee
as long as
\be 
M L \gg 1
\ee
corresponding to an induced Compton wavelength for the scalar much smaller than the scales of substructures $L$. This is the screening criterion guaranteeing that all substructures are screened. The fact that the scalar perturbations do not propagate further than the small Compton wavelength $M^{-1}$ is reminiscent of the penetration of magnetic field no further than the London length in superconductors and corresponds to a cosmic Meissner effect, see appendix (\ref{app:super}) for more detail. . 

\subsection{The noise at leading order}

The noise plays a significant role in the perturbations on large scales and acts as a source term for long-wavelength fluctuations (\ref{pert1},\ref{pert2}). 
The correlation functions of the noise $\xi$ are given by the cumulants of even order $C_{2p}(\Gamma)$ expanded in powers of the field $\phi_-$. The leading order correlation will be the two-point function, as if the noise were Gaussian. The higher order correlations are suppressed by powers of $\beta/m_{\rm Pl}$. 
At leading order we have 
\be 
\Gamma -\langle \Gamma \rangle \simeq \int d^4 x \sqrt{-g} \Delta (V(\bar\Phi+\phi_++\frac{\phi_-}{2})-V(\bar\Phi+\phi_+-\frac{\phi_-}{2})+\bar J (B(\bar\Phi+\phi_++\frac{\phi_-}{2})-B(\bar\Phi+\phi_+-\frac{\phi_-}{2})))
\ee
where we define here $\Delta A= A- \langle A\rangle$
with a leading term in $\phi_-$
\be 
\Gamma -\langle \Gamma \rangle \simeq \frac{\beta}{m_{\rm Pl}}\int d^4 x \sqrt{-g} \Delta\left [( V_1(\phi_++\bar\Phi)+ \bar J B_1(\bar \Phi+\phi_+)\right ] \phi_-.
\ee
This will give rise to the two-point function for the noise as the second cumulant is proportional to the square of $\phi_-$.

\subsubsection{Unscreened dynamics}

In this case we have to leading order
\be 
\Gamma -\langle \Gamma \rangle \simeq \frac{\beta}{m_{\rm Pl}}\int d^4 x  \sqrt{-g}j_>B_1(\phi_+) \phi_-.
\ee
This implies that the two-point function of the noise term is 
\be 
\frac{C_2(\Gamma)}{2}
\simeq \frac{\beta^2}{2 m_{\rm Pl}^2} \int d^4x d^4 y \sqrt{-g}_x\sqrt{-g}_yB_1(\phi_+(x))B_1(\phi_+(y)) \langle j_>(x) j_>(y)\rangle  \phi_-(x) \phi_-(y)
\ee
leading to the two point function
\be 
\langle \xi(x) \xi(y)\rangle \simeq \frac{\beta^2}{ m_{\rm Pl}^2} B_1(\phi_+(x))B_1(\phi_+(y)) \langle j_>(x) j_>(y)\rangle .
\ee
As we have expanded the field around the cosmological background $\phi_+= \bar \phi + \delta \phi$ in the cosmological equation, we find that the two point-function for the noise in the perturbation equation for $\delta \phi$ is identified with 
\be 
\langle \xi(\vec x,t) \xi(\vec y,t')\rangle \simeq 
\frac{\beta^2}{ m_{\rm Pl}^2}  B_1(\bar \phi(t))B_1(\bar \phi(t')) \langle j_>(\vec x, t) j_>(\vec y,t')\rangle .
\ee
The two-point function of the random field $j_>$ is given by
\be 
\langle j_>(\vec x, t) j_>(\vec y,t')\rangle= P_>(\vec x-\vec y, t,t')
\ee
where
\be 
P_>(\vec x)= \int_{\vert \vec k \vert \ge \frac{1}{L}} \slashed{d}^3 k e^{i\vec k.\vec x} P(\vec k, t,t')
\ee
is the power spectrum on short scales.
At this order we can then use the equivalence between Gaussian noises
\be 
\xi (x) \equiv \frac{\beta}{m_{\rm Pl}} B_1(\bar \phi(t)) j_>(x)
\ee
relating the noise to the short distance random source. Not surprisingly, we retrieve  
the result of linear response theory.   
Indeed the perturbative equation is now 
\be 
\delta \phi= \frac{\frac{\beta}{m_{\rm Pl}} B_1(\bar \phi) j (\vec k,\omega)}{\omega^2-\vec k^2 -M^2} 1_{k\le \frac{1}{L}}
\ee
where $j= \delta j_< +j_>$ and $j_<= \bar J_0 + \delta j_<$.
We have obtained that the cosmological perturbations respond to the total matter perturbations. As this equation is projected on scales $k\le \frac{1}{L}$  the contribution from the noise $\xi$  cancels, i.e. it only plays a role on short scales. There is effectively no effect from the openness of the system on large scales.  The only contribution  of the small scales on the large scale dynamics is the small contribution in $M^2$. 

\subsubsection{Screened dynamics}
In this case the leading order term is 
\be 
\Gamma -\langle \Gamma \rangle \simeq \frac{\beta}{m_{\rm Pl}}\int d^4 x \sqrt{-g}\Delta\left [( V_1(\Phi)+ \bar J B_1(\Phi)\right ] \phi_-.
\ee
Using the minimum equation we have 
\be 
\Gamma -\langle \Gamma \rangle \simeq \frac{\beta}{m_{\rm Pl}}\int d^4 x  \sqrt{-g} J_0 \Delta (B_1(\Phi))  \phi_-
\ee
implying that
\be 
\langle \xi (x) \xi (y) \rangle \simeq (\frac{\beta}{m_{\rm Pl}})^2 \bar J_0 (t) \bar J_0 (t') \langle \Delta B_1(\bar \Phi)(x) \Delta B_1 (\bar \Phi)(y)\rangle 
\ee
As an order of magnitude we have  $\xi ={\cal O}( \beta \bar J_0/m_{\rm Pl})$. In (\ref{pert2}) the noise contribution is potentially larger than the source coming from the large scale fluctuations $\delta j_<$ contrary to the unscreened case. As a result, we have for the perturbations on large scales
\be  
\frac{\beta\delta \phi}{m_{\rm Pl}}
\simeq \beta^2 \frac{H^2}{M^2}
\ee
which is small as  $M\gg H$. The dynamics on large scale are frozen. This is similar to the chameleon mechanism applied to the whole cosmic web where all structures are screened and therefore light scalar fields undergo the equivalent of a cosmic Meissner effect.

\section{conclusion}

We have studied the effects of small scale structures on the large scale dynamics of very light fields. We have focused on the test case where the dynamics of the Universe are not influenced by the scalar fields. Hence the matter distribution viewed as a stochastic source is the one of the cosmological standard model. In this cosmic web, the light scalars interact with structures on all scales and have effective interactions resulting from the entanglement with matter. We have used an approach very common in particle physics and integrated out the dynamics on short scales.  The very same effective field theory method is also applied successfully for the emission of gravitational and scalar waves \cite{Porto:2024cwd}. On large scales, this procedure leads to new effective interactions between light fields which influence the classical evolution and here the scalar Klein-Gordon equation. The case involving cosmic structure has a twist as the matter distribution is random. For a given realisation, the statistical properties of the system can be attained by ergodicity and averaging over the Universe. This is what we do and we obtain a grand potential after averaging over the matter fluctuation. This grand potential contains random sources characteristic of an open system. They are intrinsic and reflect the exchange of information between short and large scales. 
In a cosmological setting, the background evolution of the scalars on large scales is independent of the noise and corrected by new effective interactions induced from small scales. The noise plays a role only at the perturbative level. 

In practice, we have considered two extreme situations. The first one is the unscreened case corresponding to the weak coupling limit. The dynamics on large scales are corrected by operators appearing in a power series in the weak coupling. Small perturbations to the background cosmology follow the linear response theory to the matter fluctuations. On the other hand, when all substructures are screened then the large scale evolution of the light scalars is extremely limited. Essentially propagation is cut at the small Compton wavelength induced by the short distance dynamics.
Of course, the behaviour of light dark energy fields coupled to matter would not automatically conform to these extreme cases. Indeed the growth of structure is influenced by the dynamics of dark energy which then could be influenced by the structures themselves. This bootstrapping and back-reaction problem is beyond the scope of the present paper. All the more as screening would not be efficient on all scales, leaving unscreened regions of the Universe interspersed with screened ones. It is to be expected that the dynamics of light dark energy fields could be influenced by the present of substructures especially in a mixed scenario where solar and subgalactic scales would be screened and larger structures such as galaxies potentially unscreened. In this case, the coupling of the light scalar fields  to dark matter on galactic scales could lead to the emergence of an effective dark energy \cite{Contaldi:2026vsw} on large scales. This would provide a solution to the coincidence problem as hinted in section \ref{sec:bac}. As cosmology is reaching the percent level of accuracy, one should quantify the back-reaction discussed in this paper and check if its influence should not be taken into account by precision cosmology. This may require large scale computer simulations and is beyond what is presented here. 
\acknowledgments
I would like to thank C. Burgess for suggesting the possible influence of small scale matter fluctuations on large scale scalar dynamics. C. van de Bruck, M. Mylova and A. smith shared early discussions on this project. T. Colas, A. Davis and F. Vernizzi provided useful comments on the manuscript. 

\appendix
\section{Green's functions and classical solutions}
\label{app:green}
In the Schwinger-Keldysh formalism, the Green's functions satisfy
\begin{equation}
    (\square - m^2) G_{ab} = \delta^{(4)}(x-y)c_{ab}
\end{equation}
with $c_{11}=1$, $c_{22}=-1$ and $c_{12}=c_{21}=0$.
Here $G_{11}=G_F$ is the Feynman Green's function, $G_{22}=-G_F^*$,  $G_{12}$  is the negative frequency part whilst $G_{21}$ is the positive frequency part. Explicitly and in Fourier space we have
\begin{equation}
    \left\lbrace \begin{array}{ll}
	G_{11}= \frac{1}{\omega^2-\vec p^2-m^2+i\epsilon} & G_{12}= -i\theta(-\omega)\slashed{\delta}(\omega^2-\vec p^2-m^2) \\
	G_{21}= -i\theta(\omega)\slashed{\delta}(\omega^2-\vec p^2-m^2) & G_{22}= \frac{-1}{\omega^2-\vec p^2-m^2-i\epsilon}
\end{array}\right.
\end{equation}
The functions $\Delta=iG$ are the usual quantum field theory propagators. Notice that $G_H= \frac{G_{11}+G_{22}}{2}= \frac{G_{12}+G_{21}}{2}$ is the Hadamard Green's function involving only on-shell propagation.
In the $\pm$ basis we have the matrix of propagators $G^{ab}= c^{ac} c^{bd} G_{cd}$ and 
\begin{equation}
	G^{ab}= 	\begin{pmatrix}
0 & G_{\rm adv}\\
G_{\rm ret} & G_H
\end{pmatrix}.
	\end{equation}
where $G_{\rm ret}$ is the retarded Green's function and $G_{\rm adv}$ the advanced one. 
This implies  that classical equations of the type
\be 
(\Box -m^2)\phi_a= J_a
\ee
where $\phi_a=(\phi_+,\phi_-)$ and $J_a= (J_+, J_-)$ are solved by
\be 
\phi^a (x)= (G^{ab}\star J_b)(x)
\ee
with summation over the doublet indices and 
$J^a= c^{ab}J_b$ with $c^{ab} c_{bc}= \delta^{a}_c$. In the $\pm$ basis we have $c_{+-}=c_{-+}=1$ and similarly $c^{+-}=c^{-+}=1$. Indices are raised and lowered with $c_{ab}$ and $c^{ab}$.
In particular, the classical field is
\be 
\phi_+(x)= \phi^-(x)= (G^{-+}\star J_++ G^{--}\star J_-)(x)
\ee
where $J_-=0$  and $J_+=\bar J$ for classical configurations implying that
\be 
\phi_+ (x) = (G_{\rm ret}\star \bar J)(x)
\ee
as used in the main text. As can be seen, using the $\pm$ basis simplifies the coupling between the two Keldysh contours and makes the classical evolution similar to the one in a classical radiation problem where the retarded Green's function is used.

\section{An electromagnetic analogy}
\label{app:super}
Matter can have strong effects on the propagation of light fields. This is what happens for photons travelling at the speed of light and affected by the interactions with charged particles in a medium. This can lead to the slowing down of propagation like in glass or complete opacity in the case of superconductors where static magnetic fields are expelled from a superconducting slab \cite{kittel}. 
These behaviours can be understood by writing the propagation equation in Fourier space with a non-trivial permittivity function $\epsilon(\omega)$
\be
(  |\vec k|^2 - \epsilon (\omega) \, \omega^2) \vec A=0 \, ,
\ee
where $\vec A$ is the vector potential and Maxwell's equations have been written in the Coulomb gauge $A_0=0$. In the most simplistic model of matter, the Drude model, where the free carriers are the electrons, we have
\be 
\epsilon(\omega)= 1- \frac{\omega^2_{\rm Pl}}{\omega^2} \, ,
\ee
where
\be 
\omega_{\rm Pl}^2= \frac{16\pi^2 \alpha n}{m_e} \, ,
\ee
is the plasma frequency and  $\alpha=e^2/4\pi$  the fine structure constant, $n$ the density of carriers and $m_e$ the electron mass. This models applies to a plasma and to a superconductor where the resistivity vanishes. 
The effective dispersion relation is
\be 
\omega (k)= \sqrt{ \vec k^2 + \omega^2_{\rm Pl}} \, .
\ee
At high energy above the plasma frequency, the medium is transparent and waves have a   group velocity
\be 
v_{\rm gr}= \frac{d\omega(k)}{dk}= \frac{k}{\omega (k)}\le 1 \, ,
\ee
which is always less than unity.
In this case, propagation is allowed at a reduced speed.
On the contrary, at low energy $\omega \le \omega_{\rm Pl}$, propagation is not allowed and static fields do not penetrate further than the London length 
\be 
\lambda_{L}= \frac{1}{\omega_{\rm Pl}} \, .
\ee
This is what happens to a static magnetic field satisfying the London equation
\be 
\Delta \vec B= \omega_{\rm Pl}^2 \vec B \, ,
\ee
leading to the Meissner effect.
Of course, all this is simply a manifestation that in matter the photons acquire a mass 
\be 
m_\gamma= \omega_{\rm Pl} \, .
\ee
Viewed in this fashion, the Meissner effect for low frequency photons is nothing but screening of the photon field by matter on a distance scale corresponding to the Compton wavelength of photons. 
In the absence of screening at high enough frequency, the photon field simply travels at a speed less than the speed of light.

In the main text, we are interested in similar phenomena for light scalar fields.  We consider that they interact with matter on cosmological scales and also locally with the large scale structures of the Universe. The latter matter density can be large in galaxies for instance and is viewed as a stochastic distribution on short scales influencing the propagation and evolution of the dark energy field on large scales. On such large scales, the background evolution is modified by the presence of matter. Cosmological perturbations of the light field are also affected. We  find that complete screening of the scalar field simply prevents any dynamical evolution  whilst small corrections to the cosmological dynamics apply in the unscreened case. Cosmological perturbations have the same type of features as photons do in matter, i.e. complete screening prevents  any propagation of light  fields whilst in the unscreened case the light fields have a modified velocity coming from a matter-induced contribution to the scalar's mass.

\bibliography{ref}

\begin{thebibliography}{44}
\expandafter\ifx\csname natexlab\endcsname\relax\def\natexlab#1{#1}\fi
\expandafter\ifx\csname bibnamefont\endcsname\relax
  \def\bibnamefont#1{#1}\fi
\expandafter\ifx\csname bibfnamefont\endcsname\relax
  \def\bibfnamefont#1{#1}\fi
\expandafter\ifx\csname citenamefont\endcsname\relax
  \def\citenamefont#1{#1}\fi
\expandafter\ifx\csname url\endcsname\relax
  \def\url#1{\texttt{#1}}\fi
\expandafter\ifx\csname urlprefix\endcsname\relax\def\urlprefix{URL }\fi
\providecommand{\bibinfo}[2]{#2}
\providecommand{\eprint}[2][]{\url{#2}}

\bibitem[{\citenamefont{Copeland et~al.}(2006)\citenamefont{Copeland, Sami, and
  Tsujikawa}}]{Copeland:2006wr}
\bibinfo{author}{\bibfnamefont{E.~J.} \bibnamefont{Copeland}},
  \bibinfo{author}{\bibfnamefont{M.}~\bibnamefont{Sami}}, \bibnamefont{and}
  \bibinfo{author}{\bibfnamefont{S.}~\bibnamefont{Tsujikawa}},
  \bibinfo{journal}{Int. J. Mod. Phys. D} \textbf{\bibinfo{volume}{15}},
  \bibinfo{pages}{1753} (\bibinfo{year}{2006}), \eprint{hep-th/0603057}.

\bibitem[{\citenamefont{Joyce et~al.}(2015)\citenamefont{Joyce, Jain, Khoury,
  and Trodden}}]{Joyce:2014kja}
\bibinfo{author}{\bibfnamefont{A.}~\bibnamefont{Joyce}},
  \bibinfo{author}{\bibfnamefont{B.}~\bibnamefont{Jain}},
  \bibinfo{author}{\bibfnamefont{J.}~\bibnamefont{Khoury}}, \bibnamefont{and}
  \bibinfo{author}{\bibfnamefont{M.}~\bibnamefont{Trodden}},
  \bibinfo{journal}{Phys. Rept.} \textbf{\bibinfo{volume}{568}},
  \bibinfo{pages}{1} (\bibinfo{year}{2015}), \eprint{1407.0059}.

\bibitem[{\citenamefont{Brax}(2018)}]{Brax:2017idh}
\bibinfo{author}{\bibfnamefont{P.}~\bibnamefont{Brax}}, \bibinfo{journal}{Rept.
  Prog. Phys.} \textbf{\bibinfo{volume}{81}}, \bibinfo{pages}{016902}
  (\bibinfo{year}{2018}).

\bibitem[{\citenamefont{Adame et~al.}(2024)}]{DESI:2024mwx}
\bibinfo{author}{\bibfnamefont{A.~G.} \bibnamefont{Adame}} \bibnamefont{et~al.}
  (\bibinfo{collaboration}{DESI}) (\bibinfo{year}{2024}), \eprint{2404.03002}.

\bibitem[{\citenamefont{Lodha et~al.}(2025)}]{DESI:2025fii}
\bibinfo{author}{\bibfnamefont{K.}~\bibnamefont{Lodha}} \bibnamefont{et~al.}
  (\bibinfo{collaboration}{DESI}) (\bibinfo{year}{2025}), \eprint{2503.14743}.

\bibitem[{\citenamefont{Abdul~Karim et~al.}(2025)}]{DESI:2025zgx}
\bibinfo{author}{\bibfnamefont{M.}~\bibnamefont{Abdul~Karim}}
  \bibnamefont{et~al.} (\bibinfo{collaboration}{DESI}) (\bibinfo{year}{2025}),
  \eprint{2503.14738}.

\bibitem[{\citenamefont{Abbott et~al.}(2026)}]{DES:2026jmi}
\bibinfo{author}{\bibfnamefont{T.~M.~C.} \bibnamefont{Abbott}}
  \bibnamefont{et~al.} (\bibinfo{collaboration}{DES}) (\bibinfo{year}{2026}),
  \eprint{2605.27221}.

\bibitem[{\citenamefont{Amendola}(2000)}]{Amendola:1999er}
\bibinfo{author}{\bibfnamefont{L.}~\bibnamefont{Amendola}},
  \bibinfo{journal}{Phys. Rev. D} \textbf{\bibinfo{volume}{62}},
  \bibinfo{pages}{043511} (\bibinfo{year}{2000}), \eprint{astro-ph/9908023}.

\bibitem[{\citenamefont{Brax}(2025)}]{Brax:2025ahm}
\bibinfo{author}{\bibfnamefont{P.}~\bibnamefont{Brax}}, \bibinfo{journal}{Phys.
  Rev. D} \textbf{\bibinfo{volume}{112}}, \bibinfo{pages}{083544}
  (\bibinfo{year}{2025}), \eprint{2507.16723}.

\bibitem[{\citenamefont{Bertotti et~al.}(2003)\citenamefont{Bertotti, Iess, and
  Tortora}}]{Bertotti:2003rm}
\bibinfo{author}{\bibfnamefont{B.}~\bibnamefont{Bertotti}},
  \bibinfo{author}{\bibfnamefont{L.}~\bibnamefont{Iess}}, \bibnamefont{and}
  \bibinfo{author}{\bibfnamefont{P.}~\bibnamefont{Tortora}},
  \bibinfo{journal}{Nature} \textbf{\bibinfo{volume}{425}},
  \bibinfo{pages}{374} (\bibinfo{year}{2003}).

\bibitem[{\citenamefont{Archidiacono et~al.}(2022)\citenamefont{Archidiacono,
  Castorina, Redigolo, and Salvioni}}]{Archidiacono:2022iuu}
\bibinfo{author}{\bibfnamefont{M.}~\bibnamefont{Archidiacono}},
  \bibinfo{author}{\bibfnamefont{E.}~\bibnamefont{Castorina}},
  \bibinfo{author}{\bibfnamefont{D.}~\bibnamefont{Redigolo}}, \bibnamefont{and}
  \bibinfo{author}{\bibfnamefont{E.}~\bibnamefont{Salvioni}},
  \bibinfo{journal}{JCAP} \textbf{\bibinfo{volume}{10}}, \bibinfo{pages}{074}
  (\bibinfo{year}{2022}), \eprint{2204.08484}.

\bibitem[{\citenamefont{G{\'o}mez-Valent
  et~al.}(2020)\citenamefont{G{\'o}mez-Valent, Pettorino, and
  Amendola}}]{Gomez-Valent:2020mqn}
\bibinfo{author}{\bibfnamefont{A.}~\bibnamefont{G{\'o}mez-Valent}},
  \bibinfo{author}{\bibfnamefont{V.}~\bibnamefont{Pettorino}},
  \bibnamefont{and} \bibinfo{author}{\bibfnamefont{L.}~\bibnamefont{Amendola}},
  \bibinfo{journal}{Phys. Rev. D} \textbf{\bibinfo{volume}{101}},
  \bibinfo{pages}{123513} (\bibinfo{year}{2020}), \eprint{2004.00610}.

\bibitem[{\citenamefont{Das et~al.}(2006)\citenamefont{Das, Corasaniti, and
  Khoury}}]{Das:2005yj}
\bibinfo{author}{\bibfnamefont{S.}~\bibnamefont{Das}},
  \bibinfo{author}{\bibfnamefont{P.~S.} \bibnamefont{Corasaniti}},
  \bibnamefont{and} \bibinfo{author}{\bibfnamefont{J.}~\bibnamefont{Khoury}},
  \bibinfo{journal}{Phys. Rev. D} \textbf{\bibinfo{volume}{73}},
  \bibinfo{pages}{083509} (\bibinfo{year}{2006}), \eprint{astro-ph/0510628}.

\bibitem[{\citenamefont{Wolf et~al.}(2025)\citenamefont{Wolf, Ferreira, and
  Garc{\'\i}a-Garc{\'\i}a}}]{Wolf:2024stt}
\bibinfo{author}{\bibfnamefont{W.~J.} \bibnamefont{Wolf}},
  \bibinfo{author}{\bibfnamefont{P.~G.} \bibnamefont{Ferreira}},
  \bibnamefont{and}
  \bibinfo{author}{\bibfnamefont{C.}~\bibnamefont{Garc{\'\i}a-Garc{\'\i}a}},
  \bibinfo{journal}{Phys. Rev. D} \textbf{\bibinfo{volume}{111}},
  \bibinfo{pages}{L041303} (\bibinfo{year}{2025}), \eprint{2409.17019}.

\bibitem[{\citenamefont{Khoury et~al.}(2025)\citenamefont{Khoury, Lin, and
  Trodden}}]{Khoury:2025txd}
\bibinfo{author}{\bibfnamefont{J.}~\bibnamefont{Khoury}},
  \bibinfo{author}{\bibfnamefont{M.-X.} \bibnamefont{Lin}}, \bibnamefont{and}
  \bibinfo{author}{\bibfnamefont{M.}~\bibnamefont{Trodden}},
  \bibinfo{journal}{Phys. Rev. Lett.} \textbf{\bibinfo{volume}{135}},
  \bibinfo{pages}{181001} (\bibinfo{year}{2025}), \eprint{2503.16415}.

\bibitem[{\citenamefont{Martin et~al.}(2006)\citenamefont{Martin, Schimd, and
  Uzan}}]{Martin:2005bp}
\bibinfo{author}{\bibfnamefont{J.}~\bibnamefont{Martin}},
  \bibinfo{author}{\bibfnamefont{C.}~\bibnamefont{Schimd}}, \bibnamefont{and}
  \bibinfo{author}{\bibfnamefont{J.-P.} \bibnamefont{Uzan}},
  \bibinfo{journal}{Phys. Rev. Lett.} \textbf{\bibinfo{volume}{96}},
  \bibinfo{pages}{061303} (\bibinfo{year}{2006}), \eprint{astro-ph/0510208}.

\bibitem[{\citenamefont{Brax et~al.}(2021)\citenamefont{Brax, Casas, Desmond,
  and Elder}}]{Brax:2021wcv}
\bibinfo{author}{\bibfnamefont{P.}~\bibnamefont{Brax}},
  \bibinfo{author}{\bibfnamefont{S.}~\bibnamefont{Casas}},
  \bibinfo{author}{\bibfnamefont{H.}~\bibnamefont{Desmond}}, \bibnamefont{and}
  \bibinfo{author}{\bibfnamefont{B.}~\bibnamefont{Elder}},
  \bibinfo{journal}{Universe} \textbf{\bibinfo{volume}{8}}, \bibinfo{pages}{11}
  (\bibinfo{year}{2021}), \eprint{2201.10817}.

\bibitem[{\citenamefont{Burrage and Sakstein}(2018)}]{Burrage:2017qrf}
\bibinfo{author}{\bibfnamefont{C.}~\bibnamefont{Burrage}} \bibnamefont{and}
  \bibinfo{author}{\bibfnamefont{J.}~\bibnamefont{Sakstein}},
  \bibinfo{journal}{Living Rev. Rel.} \textbf{\bibinfo{volume}{21}},
  \bibinfo{pages}{1} (\bibinfo{year}{2018}), \eprint{1709.09071}.

\bibitem[{\citenamefont{Khoury and
  Weltman}(2004{\natexlab{a}})}]{Khoury:2003aq}
\bibinfo{author}{\bibfnamefont{J.}~\bibnamefont{Khoury}} \bibnamefont{and}
  \bibinfo{author}{\bibfnamefont{A.}~\bibnamefont{Weltman}},
  \bibinfo{journal}{Phys. Rev. Lett.} \textbf{\bibinfo{volume}{93}},
  \bibinfo{pages}{171104} (\bibinfo{year}{2004}{\natexlab{a}}),
  \eprint{astro-ph/0309300}.

\bibitem[{\citenamefont{Khoury and
  Weltman}(2004{\natexlab{b}})}]{Khoury:2003rn}
\bibinfo{author}{\bibfnamefont{J.}~\bibnamefont{Khoury}} \bibnamefont{and}
  \bibinfo{author}{\bibfnamefont{A.}~\bibnamefont{Weltman}},
  \bibinfo{journal}{Phys. Rev. D} \textbf{\bibinfo{volume}{69}},
  \bibinfo{pages}{044026} (\bibinfo{year}{2004}{\natexlab{b}}),
  \eprint{astro-ph/0309411}.

\bibitem[{\citenamefont{Hinterbichler and Khoury}(2010)}]{Hinterbichler:2010es}
\bibinfo{author}{\bibfnamefont{K.}~\bibnamefont{Hinterbichler}}
  \bibnamefont{and} \bibinfo{author}{\bibfnamefont{J.}~\bibnamefont{Khoury}},
  \bibinfo{journal}{Phys. Rev. Lett.} \textbf{\bibinfo{volume}{104}},
  \bibinfo{pages}{231301} (\bibinfo{year}{2010}), \eprint{1001.4525}.

\bibitem[{\citenamefont{Brax et~al.}(2010)\citenamefont{Brax, van~de Bruck,
  Davis, and Shaw}}]{Brax:2010gi}
\bibinfo{author}{\bibfnamefont{P.}~\bibnamefont{Brax}},
  \bibinfo{author}{\bibfnamefont{C.}~\bibnamefont{van~de Bruck}},
  \bibinfo{author}{\bibfnamefont{A.-C.} \bibnamefont{Davis}}, \bibnamefont{and}
  \bibinfo{author}{\bibfnamefont{D.}~\bibnamefont{Shaw}},
  \bibinfo{journal}{Phys. Rev. D} \textbf{\bibinfo{volume}{82}},
  \bibinfo{pages}{063519} (\bibinfo{year}{2010}), \eprint{1005.3735}.

\bibitem[{\citenamefont{Mota and Shaw}(2006)}]{Mota:2006ed}
\bibinfo{author}{\bibfnamefont{D.~F.} \bibnamefont{Mota}} \bibnamefont{and}
  \bibinfo{author}{\bibfnamefont{D.~J.} \bibnamefont{Shaw}},
  \bibinfo{journal}{Phys. Rev. Lett.} \textbf{\bibinfo{volume}{97}},
  \bibinfo{pages}{151102} (\bibinfo{year}{2006}), \eprint{hep-ph/0606204}.

\bibitem[{\citenamefont{Burgess}(2020)}]{Burgess:2020tbq}
\bibinfo{author}{\bibfnamefont{C.~P.} \bibnamefont{Burgess}},
  \emph{\bibinfo{title}{{Introduction to Effective Field Theory}}}
  (\bibinfo{publisher}{Cambridge University Press}, \bibinfo{year}{2020}), ISBN
  \bibinfo{isbn}{978-1-139-04804-0, 978-0-521-19547-8}.

\bibitem[{\citenamefont{Huterer}(2023)}]{Huterer:2022dds}
\bibinfo{author}{\bibfnamefont{D.}~\bibnamefont{Huterer}},
  \bibinfo{journal}{Astron. Astrophys. Rev.} \textbf{\bibinfo{volume}{31}},
  \bibinfo{pages}{2} (\bibinfo{year}{2023}), \eprint{2212.05003}.

\bibitem[{\citenamefont{Schwinger}(1961)}]{Schwinger:1960qe}
\bibinfo{author}{\bibfnamefont{J.~S.} \bibnamefont{Schwinger}},
  \bibinfo{journal}{J. Math. Phys.} \textbf{\bibinfo{volume}{2}},
  \bibinfo{pages}{407} (\bibinfo{year}{1961}).

\bibitem[{\citenamefont{Keldysh}(1964)}]{Keldysh:1964ud}
\bibinfo{author}{\bibfnamefont{L.~V.} \bibnamefont{Keldysh}},
  \bibinfo{journal}{Zh. Eksp. Teor. Fiz.} \textbf{\bibinfo{volume}{47}},
  \bibinfo{pages}{1515} (\bibinfo{year}{1964}).

\bibitem[{\citenamefont{Colas et~al.}(2025)\citenamefont{Colas, Qin, and
  Tong}}]{Colas:2025ind}
\bibinfo{author}{\bibfnamefont{T.}~\bibnamefont{Colas}},
  \bibinfo{author}{\bibfnamefont{Z.}~\bibnamefont{Qin}}, \bibnamefont{and}
  \bibinfo{author}{\bibfnamefont{X.}~\bibnamefont{Tong}}
  (\bibinfo{year}{2025}), \eprint{2512.07941}.

\bibitem[{\citenamefont{Contaldi and Pieroni}(2026)}]{Contaldi:2026vsw}
\bibinfo{author}{\bibfnamefont{C.~R.} \bibnamefont{Contaldi}} \bibnamefont{and}
  \bibinfo{author}{\bibfnamefont{M.}~\bibnamefont{Pieroni}}
  (\bibinfo{year}{2026}), \eprint{2603.23473}.

\bibitem[{\citenamefont{Shao et~al.}(2019)\citenamefont{Shao, Li, Cautun, Wang,
  and Wang}}]{Shao:2019wit}
\bibinfo{author}{\bibfnamefont{S.}~\bibnamefont{Shao}},
  \bibinfo{author}{\bibfnamefont{B.}~\bibnamefont{Li}},
  \bibinfo{author}{\bibfnamefont{M.}~\bibnamefont{Cautun}},
  \bibinfo{author}{\bibfnamefont{H.}~\bibnamefont{Wang}}, \bibnamefont{and}
  \bibinfo{author}{\bibfnamefont{J.}~\bibnamefont{Wang}},
  \bibinfo{journal}{Mon. Not. Roy. Astron. Soc.}
  \textbf{\bibinfo{volume}{489}}, \bibinfo{pages}{4912} (\bibinfo{year}{2019}),
  \eprint{1907.02081}.

\bibitem[{\citenamefont{Landau and Lifschitz}(1969)}]{LL}
\bibinfo{author}{\bibfnamefont{L.~D.} \bibnamefont{Landau}} \bibnamefont{and}
  \bibinfo{author}{\bibfnamefont{E.~M.} \bibnamefont{Lifschitz}},
  \emph{\bibinfo{title}{{Statistical Physics}}} (\bibinfo{publisher}{Pergamon
  press}, \bibinfo{year}{1969}).

\bibitem[{\citenamefont{Asgari et~al.}(2023)\citenamefont{Asgari, Mead, and
  Heymans}}]{Asgari:2023mej}
\bibinfo{author}{\bibfnamefont{M.}~\bibnamefont{Asgari}},
  \bibinfo{author}{\bibfnamefont{A.~J.} \bibnamefont{Mead}}, \bibnamefont{and}
  \bibinfo{author}{\bibfnamefont{C.}~\bibnamefont{Heymans}}
  (\bibinfo{year}{2023}), \eprint{2303.08752}.

\bibitem[{\citenamefont{Ansari et~al.}(2025)\citenamefont{Ansari, Banerjee,
  Jain, and Padhyegurjar}}]{Ansari:2024efj}
\bibinfo{author}{\bibfnamefont{A.}~\bibnamefont{Ansari}},
  \bibinfo{author}{\bibfnamefont{A.}~\bibnamefont{Banerjee}},
  \bibinfo{author}{\bibfnamefont{S.}~\bibnamefont{Jain}}, \bibnamefont{and}
  \bibinfo{author}{\bibfnamefont{S.}~\bibnamefont{Padhyegurjar}},
  \bibinfo{journal}{JCAP} \textbf{\bibinfo{volume}{10}}, \bibinfo{pages}{018}
  (\bibinfo{year}{2025}), \eprint{2406.17025}.

\bibitem[{\citenamefont{Donath et~al.}(2024)\citenamefont{Donath, Lewandowski,
  and Senatore}}]{Donath:2023sav}
\bibinfo{author}{\bibfnamefont{Y.}~\bibnamefont{Donath}},
  \bibinfo{author}{\bibfnamefont{M.}~\bibnamefont{Lewandowski}},
  \bibnamefont{and} \bibinfo{author}{\bibfnamefont{L.}~\bibnamefont{Senatore}},
  \bibinfo{journal}{Phys. Rev. D} \textbf{\bibinfo{volume}{109}},
  \bibinfo{pages}{123510} (\bibinfo{year}{2024}), \eprint{2307.11409}.

\bibitem[{\citenamefont{Porto et~al.}(2024)\citenamefont{Porto, Riva, and
  Yang}}]{Porto:2024cwd}
\bibinfo{author}{\bibfnamefont{R.~A.} \bibnamefont{Porto}},
  \bibinfo{author}{\bibfnamefont{M.~M.} \bibnamefont{Riva}}, \bibnamefont{and}
  \bibinfo{author}{\bibfnamefont{Z.}~\bibnamefont{Yang}}
  (\bibinfo{year}{2024}), \eprint{2409.05860}.

\bibitem[{\citenamefont{Brax and Bruy{\`e}re}(2025)}]{Brax:2024obu}
\bibinfo{author}{\bibfnamefont{P.}~\bibnamefont{Brax}} \bibnamefont{and}
  \bibinfo{author}{\bibfnamefont{E.}~\bibnamefont{Bruy{\`e}re}},
  \bibinfo{journal}{Phys. Rev. D} \textbf{\bibinfo{volume}{111}},
  \bibinfo{pages}{064076} (\bibinfo{year}{2025}), \eprint{2412.15092}.

\bibitem[{\citenamefont{Kaplanek et~al.}(2025)\citenamefont{Kaplanek, Mylova,
  and Tolley}}]{Kaplanek:2025moq}
\bibinfo{author}{\bibfnamefont{G.}~\bibnamefont{Kaplanek}},
  \bibinfo{author}{\bibfnamefont{M.}~\bibnamefont{Mylova}}, \bibnamefont{and}
  \bibinfo{author}{\bibfnamefont{A.~J.} \bibnamefont{Tolley}}
  (\bibinfo{year}{2025}), \eprint{2512.17089}.

\bibitem[{\citenamefont{Kaplanek et~al.}(2026)\citenamefont{Kaplanek, Mylova,
  and Tolley}}]{Kaplanek:2026kpp}
\bibinfo{author}{\bibfnamefont{G.}~\bibnamefont{Kaplanek}},
  \bibinfo{author}{\bibfnamefont{M.}~\bibnamefont{Mylova}}, \bibnamefont{and}
  \bibinfo{author}{\bibfnamefont{A.~J.} \bibnamefont{Tolley}}
  (\bibinfo{year}{2026}), \eprint{2604.26941}.

\bibitem[{\citenamefont{C{\'e}spedes and Colas}(2026)}]{Cespedes:2026fdp}
\bibinfo{author}{\bibfnamefont{S.}~\bibnamefont{C{\'e}spedes}}
  \bibnamefont{and} \bibinfo{author}{\bibfnamefont{T.}~\bibnamefont{Colas}}
  (\bibinfo{year}{2026}), \eprint{2605.11096}.

\bibitem[{\citenamefont{Amoruso et~al.}(2026)\citenamefont{Amoruso, Braga,
  Garoffolo, Lopez, Bartolo, and Matarrese}}]{Amoruso:2026txw}
\bibinfo{author}{\bibfnamefont{R.}~\bibnamefont{Amoruso}},
  \bibinfo{author}{\bibfnamefont{G.}~\bibnamefont{Braga}},
  \bibinfo{author}{\bibfnamefont{A.}~\bibnamefont{Garoffolo}},
  \bibinfo{author}{\bibfnamefont{F.}~\bibnamefont{Lopez}},
  \bibinfo{author}{\bibfnamefont{N.}~\bibnamefont{Bartolo}}, \bibnamefont{and}
  \bibinfo{author}{\bibfnamefont{S.}~\bibnamefont{Matarrese}}
  (\bibinfo{year}{2026}), \eprint{2604.15313}.

\bibitem[{\citenamefont{Donath and Pajer}(2024)}]{Donath:2024utn}
\bibinfo{author}{\bibfnamefont{Y.}~\bibnamefont{Donath}} \bibnamefont{and}
  \bibinfo{author}{\bibfnamefont{E.}~\bibnamefont{Pajer}},
  \bibinfo{journal}{JHEP} \textbf{\bibinfo{volume}{07}}, \bibinfo{pages}{064}
  (\bibinfo{year}{2024}), \eprint{2402.05999}.

\bibitem[{\citenamefont{Smith et~al.}(2025)\citenamefont{Smith, Brax, van~de
  Bruck, Burgess, and Davis}}]{Smith:2025grk}
\bibinfo{author}{\bibfnamefont{A.}~\bibnamefont{Smith}},
  \bibinfo{author}{\bibfnamefont{P.}~\bibnamefont{Brax}},
  \bibinfo{author}{\bibfnamefont{C.}~\bibnamefont{van~de Bruck}},
  \bibinfo{author}{\bibfnamefont{C.~P.} \bibnamefont{Burgess}},
  \bibnamefont{and} \bibinfo{author}{\bibfnamefont{A.-C.} \bibnamefont{Davis}},
  \bibinfo{journal}{Eur. Phys. J. C} \textbf{\bibinfo{volume}{85}},
  \bibinfo{pages}{1062} (\bibinfo{year}{2025}), \eprint{2505.05450}.

\bibitem[{\citenamefont{Brax et~al.}(2004)\citenamefont{Brax, van~de Bruck,
  Davis, Khoury, and Weltman}}]{Brax:2004qh}
\bibinfo{author}{\bibfnamefont{P.}~\bibnamefont{Brax}},
  \bibinfo{author}{\bibfnamefont{C.}~\bibnamefont{van~de Bruck}},
  \bibinfo{author}{\bibfnamefont{A.-C.} \bibnamefont{Davis}},
  \bibinfo{author}{\bibfnamefont{J.}~\bibnamefont{Khoury}}, \bibnamefont{and}
  \bibinfo{author}{\bibfnamefont{A.}~\bibnamefont{Weltman}},
  \bibinfo{journal}{Phys. Rev. D} \textbf{\bibinfo{volume}{70}},
  \bibinfo{pages}{123518} (\bibinfo{year}{2004}), \eprint{astro-ph/0408415}.

\bibitem[{\citenamefont{Kittel}(2004)}]{kittel}
\bibinfo{author}{\bibfnamefont{C.}~\bibnamefont{Kittel}},
  \emph{\bibinfo{title}{{Introduction to Solid State Physics}}}
  (\bibinfo{publisher}{Wiley}, \bibinfo{year}{2004}).

\end{thebibliography}

\end{document}